\documentclass{article}
\pdfoutput=1
\usepackage{amsmath,amssymb,amsfonts}
\usepackage{graphicx,color}
\usepackage{calc}
\usepackage{multicol}
\usepackage[hyperindex,breaklinks]{hyperref}
\usepackage[numeric,initials]{amsrefs}
\usepackage[width=372pt,height=588pt,top=126pt]{geometry} % IOP journal standard textwidth and text height

\newcommand{\figtobox}[3][{}]{%
	\expandafter\newsavebox\csname #2box\endcsname%
	\expandafter\savebox\csname #2box\endcsname{#1{\input{#3.pdf_t}}}%
	\expandafter\def\csname #2\endcsname%
		{\setlength{\figoff}{1.0ex-.5\expandafter\ht\csname #2box\endcsname}%
			\raisebox{\figoff}{\expandafter\usebox\csname #2box\endcsname}}%
}
\newlength{\figoff}

  \newcommand{\bj}{\mathbf{j}}
\renewcommand{\d}{\mathrm{d}}
  \newcommand{\Spin}{\mathrm{Spin}}
  
  \newcommand{\Tet}[1]{\ensuremath{\left[\begin{matrix}#1\end{matrix}\right]}}
  \newcommand{\half}{{\textstyle\frac{1}{2}}}
  \newcommand{\tr}{\operatorname{tr}}
  \newcommand{\isom}{\cong}
  
  \newcommand{\tri}{\operatorname{tri}}
\title{Evaluation of new spin foam vertex amplitudes}
\author{Igor Khavkine \\
	\small
	Department of Applied Mathematics,\\
	\small
	University of Western Ontario, London, ON N6A 5B7, Canada\\
	\small
	E-mail: ikhavkin@alumni.uwo.ca}

\begin{document}
\maketitle
%%% Figures to include
%% Figures from newvert.
\figtobox[%
		\def\j{ }%
	]{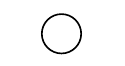}{loopsn}%
\figtobox[%
		\def\j{ }%
		\def\jp{ }%
		\def\twok{ }%
	]{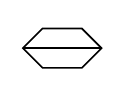}{thetasn}%
\figtobox[%
		\def\intg{$\displaystyle \int \d{g}$}%
		\def\j{$\bj^*$}%
	]{edgeint}{edge-int}%
\figtobox[%
		\def\intg{$\displaystyle \int \d{g}$}%
		\def\intgp{$\displaystyle \int \d{g'}$}%
		\def\pr{$\prime$}%
		\def\j{$\bj^*$}%
	]{edgeinttwo}{edge-int2}%
\figtobox[%
		\def\sumi{$\sum_{i}$}%
		\def\sumip{$\sum_{i'}$}%
		\def\i{$i$}%
		\def\ip{$i'$}%
		\def\j{$\bj^*$}%
	]{edgesum}{edge-sum}%
\figtobox[%
		\def\jj{$\bj$}%
	]{idprlone}{id-prl-one}%
\figtobox[%
		\def\jl{$j$}%
		\def\jr{$j$}%
	]{idprltwo}{id-prl-two}%
\figtobox[%
		\def\sumk{$\displaystyle \sum_{k=0}^j C_{jk} \frac{\loopsn}{\thetasn}$}%
		\def\twok{$2k$}%
		\def\jl{$j$}%
		\def\jr{$j$}%
	]{recprl}{rec-prl}%
\figtobox[%
		\def\sumk{$\displaystyle \sum_{k=0}^j \frac{\loopsn 2k}{\thetasn j^*,2k}$}%
		\def\twok{$2k$}%
		\def\jl{$j$}%
		\def\jr{$j$}%
	]{recprlnosum}{rec-prl}%
\figtobox[%
		\def\intg{$\displaystyle \int \d{g^-}\d{g^+}$}%
		\def\twok{$2k$}%
		\def\jl{$j$}%
		\def\jr{$j$}%
	]{gapre}{ga-pre}%
\figtobox[%
		\def\twok{$2k^*$}%
		\def\twoi{$2i$}%
		\def\il{$i^-$}%
		\def\ir{$i^+$}%
		\def\jl{$j^*$}%
		\def\jr{$j^*$}%
		\def\sumi{$\displaystyle \sum_{i,i^-,i^+}$}%
		\def\ik{$2i,2k^*$}%
		\def\ilj{$i^-,j^*$}%
		\def\irj{$i^+,j^*$}%
	]{gapost}{ga-post}%
\figtobox[%
		\def\twoi{$2i$}%
		\def\twok{$2k$}%
		\def\twokp{$2k'$}%
		\def\j{$j$}%
		\def\jp{$j'$}%
		\def\il{$i^-$}%
		\def\ir{$i^+$}%
	]{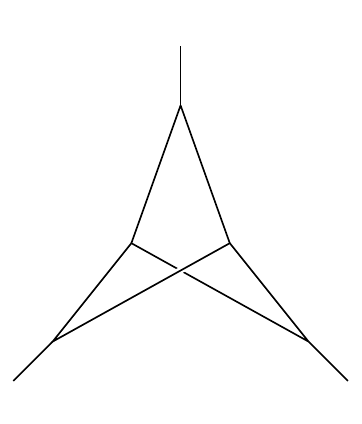}{klingon}%
\figtobox[%
		\def\twoi{$2i$}%
		\def\twok{$2k$}%
		\def\twokp{$2k'$}%
		\def\j{$j$}%
		\def\jp{$j'$}%
		\def\il{$i^-$}%
		\def\ir{$i^+$}%
		\def\n{$n$}%
	]{klingoni}{klingon-rec1}%
\figtobox[%
		\def\twoi{$2i$}%
		\def\twok{$2k$}%
		\def\twokp{$2k'$}%
		\def\il{$i^-$}%
		\def\ir{$i^+$}%
		\def\n{$n$}%
	]{klingonii}{klingon-rec2}%
\figtobox[%
		\def\twoi{$2i$}%
		\def\il{$i^-$}%
		\def\ir{$i^+$}%
	]{klingoniii}{klingon-rec3}%
\figtobox[%
		\def\twoi{$2i$}%
		\def\il{$i^-$}%
		\def\ir{$i^+$}%
	]{thetatw}{tri-p-v}%
\figtobox[%
		\def\twoi{$2i$}%
		\def\twok{$2k^*$}%
		\def\j{$j^*$}%
		\def\il{$i^-$}%
		\def\ir{$i^+$}%
	]{tripetal}{tri-p}%
\figtobox[%
		\def\twoi{$2i$}%
		\def\twok{$2k^*$}%
		\def\j{$j^*$}%
		\def\il{$i^-$}%
		\def\ir{$i^+$}%
	]{triplong}{tri-p-long}%
\figtobox[%
		\def\twoi{$2i$}%
		\def\twok{$2j^*$}%
		\def\j{$j^*$}%
		\def\il{$i^-$}%
		\def\ir{$i^+$}%
	]{tripetalepr}{tri-p}%
\figtobox[%
		\def\ip{$i'$}%
		\def\i{$i$}%
		\def\j{$\bj_1$}%
		\def\k{$\bj_2$}%
		\def\l{$\bj_3$}%
		\def\m{$\bj_4$}%
	]{edgenorm}{edge-norm}%
\figtobox[%
		\def\a{$i_0$}%
		\def\b{$i_1$}%
		\def\c{$i_2$}%
		\def\d{$i_3$}%
		\def\e{$i_4$}%
		\def\joa{$\bj_{1,0}$}%
		\def\job{$\bj_{1,1}$}%
		\def\joc{$\bj_{1,2}$}%
		\def\jod{$\bj_{1,3}$}%
		\def\joe{$\bj_{1,4}$}%
		\def\jia{$\bj_{2,0}$}%
		\def\jib{$\bj_{2,1}$}%
		\def\jic{$\bj_{2,2}$}%
		\def\jid{$\bj_{2,3}$}%
		\def\jie{$\bj_{2,4}$}%
	]{blobpent}{pent-blob}
\figtobox[%
		\def\i{}%
		\def\job{$\bj_1$}%
		\def\jib{$\bj_2$}%
		\def\jie{$\bj_3$}%
		\def\joa{$\bj_4$}%
	]{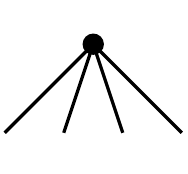}{bclhs}%
\figtobox[%
		\def\i{$i$}%
		\def\job{$j_1$}%
		\def\jib{$j_2$}%
		\def\jie{$j_3$}%
		\def\joa{$j_4$}%
	]{bclhsi}{bclhs}%
\figtobox[%
		\def\i{$i,k^*$}%
		\def\job{$\bj_1$}%
		\def\jib{$\bj_2$}%
		\def\jie{$\bj_3$}%
		\def\joa{$\bj_4$}%
	]{bclhsik}{bclhs}%
\figtobox[%
		\def\i{$i$}%
		\def\job{$j_1$}%
		\def\jib{$j_2$}%
		\def\jie{$j_3$}%
		\def\joa{$j_4$}%
	]{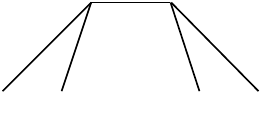}{bcrhs}%
\figtobox[%
		\def\i{$i^-$}%
		\def\job{$j_1$}%
		\def\jib{$j_2$}%
		\def\jie{$j_3$}%
		\def\joa{$j_4$}%
	]{bcrhsm}{bcrhs}%
\figtobox[%
		\def\i{$i^+$}%
		\def\job{$j_1$}%
		\def\jib{$j_2$}%
		\def\jie{$j_3$}%
		\def\joa{$j_4$}%
	]{bcrhsp}{bcrhs}%
\figtobox[%
		\def\a{}%
		\def\b{}%
		\def\c{}%
		\def\d{}%
		\def\e{}%
		\def\joa{\large ~$\bj^*$}%
		\def\job{}%
		\def\joc{}%
		\def\jod{}%
		\def\joe{}%
		\def\jia{}%
		\def\jib{}%
		\def\jic{}%
		\def\jid{}%
		\def\jie{}%
	]{tenj}{10j-cros}%
\figtobox[%
		\def\a{}%
		\def\b{\large ~$i^*$}%
		\def\c{}%
		\def\d{}%
		\def\e{}%
		\def\joa{\large ~$\bj^*$}%
		\def\job{}%
		\def\joc{}%
		\def\jod{}%
		\def\joe{}%
		\def\jia{}%
		\def\jib{}%
		\def\jic{}%
		\def\jid{}%
		\def\jie{}%
	]{tenji}{10j-cros}%
\figtobox[%
		\def\a{}%
		\def\b{\large ~$i^*,k^*$}%
		\def\c{}%
		\def\d{}%
		\def\e{}%
		\def\joa{\large ~$\bj^*$}%
		\def\job{}%
		\def\joc{}%
		\def\jod{}%
		\def\joe{}%
		\def\jia{}%
		\def\jib{}%
		\def\jic{}%
		\def\jid{}%
		\def\jie{}%
	]{tenjik}{10j-cros}%
\figtobox[%
		\def\a{}%
		\def\b{$i^*$}%
		\def\c{}%
		\def\d{}%
		\def\e{}%
		\def\joa{$j^*$}%
		\def\job{}%
		\def\joc{}%
		\def\jod{}%
		\def\joe{}%
		\def\jia{$j^*$}%
		\def\jib{}%
		\def\jic{}%
		\def\jid{}%
		\def\jie{}%
	]{fifij}{15j-cros}%
%% Figures from bdrvert.
\def\rbx#1#2{\rotatebox[origin=c]{#1}{#2}}%
\figtobox[%
		% XXX: there may be some room for improvement here
		\def\a{\rbx{288}{\rbx{-288}{$k^p$}~\rbx{-288}{$i_0$}~\rbx{-288}{$k^q$}}}%
		\def\b{\rbx{  0}{\rbx{-  0}{$k^p$}~\rbx{-  0}{$i_1$}~\rbx{-  0}{$k^q$}}}%
		\def\c{\rbx{ 72}{\rbx{- 72}{$k^p$}~\rbx{- 72}{$i_2$}~\rbx{- 72}{$k^q$}}}%
		\def\d{\rbx{144}{\rbx{-144}{$k^p$}~\rbx{-144}{$i_3$}~\rbx{-144}{$k^q$}}}%
		\def\e{\rbx{216}{\rbx{-216}{$k^p$}~\rbx{-216}{$i_4$}~\rbx{-216}{$k^q$}}}%
		\def\joa{$j_{1,0}$}%
		\def\job{$j_{1,1}$}%
		\def\joc{$j_{1,2}$}%
		\def\jod{$j_{1,3}$}%
		\def\joe{$j_{1,4}$}%
		\def\jia{$j_{2,0}$}%
		\def\jib{$j_{2,1}$}%
		\def\jic{$j_{2,2}$}%
		\def\jid{$j_{2,3}$}%
		\def\jie{$j_{2,4}$}%
	]{pentijk}{10j-cros}
\figtobox[%
		\def\T{$M_e$}%
		\def\i{$i_e$}%
		\def\ip{$i_{e+1}$}%
	]{bcM}{bc-m}%
\figtobox[%
		\def\Tl{$T^e_-$}%
		\def\Tr{$T^e_+$}%
		\def\il{$i^-_e$}%
		\def\ir{$i^+_e$}%
		\def\ilp{$i^-_{e+1}$}%
		\def\irp{$i^+_{e+1}$}%
		\def\Q{$Q^{e+1}$}%
		\def\P{$\bar{P}^e$}%
	]{newM}{new-m}%
\figtobox[%
		\def\T{$M^{\mathrm{orig}}_e$}%
		\def\p{$\psi$}%
		\def\i{$i_e$}%
		\def\ip{$i_{e+1}$}%
		\def\jt{$j_{2,e-1}$}%
		\def\jtp{$j_{2,e}$}%
		\def\jo{$j_{1,e}$}%
	]{bcbM}{bcb-m}%
\figtobox[%
		\def\A{$A_e$}%
		\def\B{$B_e$}%
		\def\C{$C_e$}%
		\def\p{$\psi$}%
		\def\i{$i_e$}%
		\def\ip{$i_{e+1}$}%
		\def\jt{$j_{2,e-1}$}%
		\def\jtp{$j_{2,e}$}%
		\def\jo{$j_{1,e}$}%
		\def\s{$s_e$}%
	]{bcbMs}{bcb-m7}%
\figtobox[%
		\def\Tl{$\bar{T}_-^e$}%
		\def\Tr{$\bar{T}_+^e$}%
		\def\p{$\psi$}%
		\def\Q{\small$Q^{e+1}$}%
		\def\P{\small$P^e$}%
		\def\i{$i_e$}%
		\def\ip{$i_{e+1}$}%
		\def\il{$i^-_e$}%
		\def\ir{$i^+_e$}%
		\def\ilp{$i^-_{e+1}$}%
		\def\irp{$i^+_{e+1}$}%
		\def\jt{$j_{2,e-1}$}%
		\def\jtp{$j_{2,e}$}%
		\def\jo{$j_{1,e}$}%
		\def\kp{\small$k^p$}%
		\def\kq{\small$k^q$}%
	]{newbM}{newb-m}%
%%% End of figures

\begin{abstract}
	\noindent
	The Christensen-Egan algorithm is extended and generalized to
	efficiently evaluate new spin foam vertex amplitudes proposed by Engle,
	Pereira \& Rovelli and Freidel \& Krasnov, with or without (factored)
	boundary states. A concrete pragmatic proposal is made for comparing the
	different models using uniform methodologies, applicable to the behavior
	of large spin asymptotics and of expectation values of specific
	semiclassical observables.  The asymptotics of the new models exhibit
	non-oscillatory, power-law decay similar to that of the Barrett-Crane
	model, though with different exponents. Also, an analysis of the
	semiclassical wave packet propagation problem indicates that the
	Magliaro, Rovelli and Perini's conjecture of good semiclassical behavior
	of the new models does not hold for generic factored states, which
	neglect spin-spin correlations.
	\vspace{1ex}

	\noindent
	PACS numbers: 04.60.Pp
\end{abstract}

\section{Introduction}
Spin foam models are an attempt to produce a theory of quantum gravity
starting from a discrete, path integral-like approach. They were first
defined a decade ago~\cites{Baez-spinfoam,BC-riem}. More recently, we
have seen significant progress toward extraction of their semiclassical
behavior and its favorable comparison to the expected weak field limit
of gravity, starting with Rovelli and collaborators' calculation of the
graviton propagator~\cites{MR,R-prop}. Unfortunately, further
calculations have revealed that the standard spin foam model due to
Barrett and Crane produced incorrect results for some of the propagator
matrix elements~\cites{AR-I,AR-II}. This result has motivated several
proposals to replace the Barrett-Crane (BC) spin foam vertex
amplitude~\cite{BC-riem} for quantum gravity. The first proposal, by
Engle, Pereira and Rovelli (EPR)~\cites{EPR,EPR-long}, aimed also to
identify the spin foam boundary state space with that of loop quantum
gravity spin networks; this model is also referred to as the ``flipped''
vertex model. Another proposal, by Livine and
Speziale~\cites{LS-coh,LS-sol}, used $SU(2)$-coherent states to define
the spin foam amplitudes and reproduced the EPR proposal up to an edge
normalization factor. Finally, a paper by Freidel and
Krasnov~\cites{FK,CF}, suggested that the EPR model corresponds to a
topological theory related to gravity and proposed a generalization
thereof corresponding to gravity itself (the FK model).  The present
paper, along with most previous work, concerns only the Riemannian
signature models of gravity.

Having been defined, the new models must be tested to see whether their
semiclassical behavior is an improvement over the BC model. The standard
tests involve examining the asymptotics of their vertex amplitudes and
checking the semiclassical behavior of observables. So far, two test
problems involving observables have been proposed: semiclassical wave
packet propagation~\cite{MPR}, and evaluation of the graviton $2$-point
function~\cites{R-prop,BMRS,CLS}. Both problems require the computation
of large sums, where the spin foam vertex amplitude is contracted with a
suitably defined boundary state. These computations, while important for
extracting the physical content of the new spin foam models, have so far
not been tractable.

This paper describes efficient numerical algorithms, based on the
existing Chris\-ten\-sen-Egan (CE) algorithm for the BC model, to evaluate
the new spin foam vertex amplitudes, both with fixed boundary spins and
contracted with a boundary state. The accessible boundary states are
restricted to the large (though with important limitations) class of
so-called \emph{factored} boundary states.  Calculation of the new
vertex amplitude asymptotics shows that they are dominated by the
degenerate spin foam sector, as is the BC model. Also, the numerical
simulation of semiclassical wave packet propagation shows that, for
factored boundary states, the shape of the propagated wave packet does
not agree with the desired semiclassical result. The state of the good
semiclassical behavior conjecture~\cite{MPR} for states that do not
neglect spin-spin correlations (unlike factored states) remains open.

Section~\ref{bf-th} briefly introduces spin foam models and presents the
three models described above in a unified framework, also incorporating
spin foams with boundary. Section~\ref{obsv-asym} describes the proposed
calculations of observables and asymptotics. The class of factored
boundary states is tentatively defined in section~\ref{sf-bdr} and in
full detail in section~\ref{num-alg}. In section~\ref{new-eval},
explicit formulas are given for new spin networks arising in the
evaluation of spin foam model vertex amplitudes. Section~\ref{num-alg}
presents efficient numerical algorithm for computation with the new
models and section~\ref{apps} shows the results of their application to
some of the problems discussed in section~\ref{obsv-asym}. Finally,
section~\ref{concl} concludes with a discussion of the results and
future work.

\section[\textit{BF} theory and spin foams]%
	{$BF$ theory and spin foams\label{bf-th}}
The new spin foam models of gravity may be presented in a way similar to
the original BC model. Following Freidel and Krasnov~\cite{FK}, we
define them within a unified framework. See also the more recent
paper~\cite{CF}.

The starting point is $BF$ theory. It is a 4-dimensional field theory
with two fields: a gauge connection 1-form $A$ and an auxiliary 2-form
$B$. The action is given by
\begin{equation}
	S = \int \tr[B\wedge F],
\end{equation}
where $F = dA$ is the curvature of the connection.  If the gauge group
is taken to be $\Spin(4)$, the double cover of $SO(4)$ and a constraint
is imposed, ensuring simplicity%
	\footnote{\emph{Simplicity} means that there exists a 1-form $e$ such
	that $B= *(e\wedge e)$, with $*$ denoting the Hodge dual.} %
of the $B$ 2-form, this theory becomes equivalent to the Plebanski
formulation of general relativity in Riemannian
signature~\cite{BC-riem}.

$BF$ theory is in a sense topological. Particularly, its underlying
manifold may be freely changed from a smooth one to a discretized
(piecewise linear) one, as long as the topology remains the same. Moreover, for
$BF$ theory, quantization and discretization commute~\cite{CY-BF}. Spin
foam models aim to reproduce gravity by heuristically imposing
simplicity constraints on $BF$ theory after the discretization and
quantization steps have been performed \cite{Baez-BF}. The connection
between spin foams and gravity is motivated by results from loop quantum
gravity~\cite{R-book}.

Consider $BF$ theory defined on a simplicial complex (possibly with
boundary), also referred to as a \emph{triangulation}. It is convenient
to introduce the \emph{dual $2$-complex}. Each $4$-simplex is identified
with a \emph{dual vertex}, each tetrahedron is identified with a
\emph{dual edge}, and each triangle is identified with a \emph{dual
face}. Discretizing the $A$ and $B$ fields and integrating out the $B$
field, the theory's path integral yields the following
expression for its partition function:
\begin{equation}\label{BF-partf}
	Z = \text{``}\int \d{A}\,\d{B}\, e^{iS}~~\text{''}=\int \d{g} \prod_f \delta(g_f),
\end{equation}
where the connection $A$ has been replaced by group elements $g$
associated to every dual edge, and $g_f$ represents the holonomy around
a dual face. This form of the partition function manifestly shows that
only flat connections (with trivial holonomies) contribute to the $BF$
theory path integral. See~\cite{O-book} for details.

The $\delta$-functions can be expanded in terms of gauge group
characters and the group integrations can be performed at each dual
edge~\cite{O-book}. What remains is a discrete sum of the form
\begin{align}
\label{sf-amp}
	A(F) &= \prod_f A_f(F) \prod_e A_e(F) \prod_v A_v(F), \\
\label{sf-partf}
	Z &= \sum_{F} A(F),
\end{align}
where $F$ ranges over all spin foams (defined below), while $f$, $e$,
and $v$ range respectively over dual faces, dual edges, and dual
vertices. In this context, a \emph{spin foam} is a labelling of the dual
faces of the triangulation by irreducible representations of the gauge
group. These representation labels come from the character expansion
described above.  This definition of spin foams will have to be
augmented with extra labels for the purpose of introducing the new
models. Thus, discrete $BF$ theory yields a \emph{spin foam model}. In
general, a spin foam model is defined by they way it labels spin foams
and by the amplitudes it gives to them through~\eqref{sf-amp} and
specific choices for $A_f(F)$, $A_e(F)$ and $A_v(F)$. Each of the
amplitudes for dual cells may depend on its own label and the labels of
adjacent cells.

Irreducible representations of $\Spin(4)\isom SU(2)\times SU(2)$ are
labelled by a pair of integers $\bj = (j^-,j^+)$, where each $j$ is a
spin%
	\footnote{Technically, a \emph{twice-spin}, since it does not take on
	half-integral values.}, %
corresponding to an irreducible representation of $SU(2)$.  Hence forth,
all representation labels will be referred to as spins, unless otherwise
specified. Bold face letters will specifically represent $\Spin(4)$
spins.

For pure $BF$ theory, face amplitudes are determined by the character
expansion of $\delta$-functions and are given by the dimension of the
irrep $\bj_f$ labelling a given face
\begin{equation}
	A_f(F) = \dim \bj_f = (j_f^{-}+1)(j_f^{+}+1).
\end{equation}
Edge and vertex amplitudes are determined by evaluating the group
integrals in equation~\eqref{BF-partf}. The basic identities we use are
\begin{equation}\label{groupint}
\scalebox{.85}{$\displaystyle
	\edgeint \hspace{-1.5em}=\hspace{-1.2em} \edgeinttwo
	\hspace{-1.5em}=\hspace{-1.7em} \edgesum.
$}
\end{equation}
The above graphical notation requires some explanation.
See~\cite{O-book} and the Appendix of~\cite{CCK} for full details.
Briefly, a solid line represents a tensor index, while any object with
lines attached to it is a tensor, with each line representing a tensor
index.  A vertical strand with a circle represents a matrix element of a
particular representation of the gauge group. Concatenation of lines
corresponds to index contraction; particularly concatenating strands
with circles corresponds to matrix multiplication. Juxtaposition of
tensorial objects corresponds to their Kronecker (tensor) product.
These graphical representations of $SU(2)$ or $\Spin(4)$ tensor
contractions will be referred to as \emph{spin networks}. Most of the
representation labels and basis indices have been omitted for
conciseness. Instead, some spins will be marked as collective labels
with an asterisk. Their expanded meaning should be clear from context.
For instance, in~\eqref{groupint} the collective label $\bj^*$ stands
for $\bj_1$, $\bj_2$, $\bj_3$ and $\bj_4$, where each strand gets its
own spin label. The blank and primed circles convey whether it is the
group element $g$ or $g'$ that is taken in the given representation. The
horizontal dotted line represents the triangulation tetrahedron dual to
the dual edge to which the given group element is associated.

The first equality in~\eqref{groupint} follows directly from the
normalization of the Haar group measure, its invariance under
translations and the multiplicative property of representation matrix
elements. In this context, group integration is also known to produce a
projection operator onto the space of intertwiners among the
representations given by the four strands. The last equality
in~\eqref{groupint} illustrates this identity by expanding this
projector in a basis of normalized intertwiners; the bracketed spin
network provides correct normalization in the denominator of the
expression. The summation over the new intertwiner basis labels $i$ and
$i'$ make up the sums over dual edge labels, part of the sum over spin
foams in \eqref{sf-partf}.  Performing the same group integration and
intertwiner expansion over all dual edges of the triangulation, we can
read off the edge and vertex amplitudes of equation~\eqref{sf-partf}.

Thus, for discrete $BF$ theory, writing all tensor contractions in terms
of spin networks we find these amplitudes to be
\begin{equation}\label{penteq}
	A_e(F) = \frac{1}{\edgenorm}
	\quad \text{and} \quad
	A_v(F) ~~= \blobpent ~.
\end{equation}
The topology of the contraction graph corresponding to $A_v$ above
follows directly from the adjacency structure of the dual $2$-complex.
We shall refer to this graph as the \emph{pent graph}; it will appear in
the vertex amplitude definition of each spin foam model discussed later
in this section. Both the edge and vertex amplitudes, $A_e$ and $A_v$,
appear with full spin labelling. For conciseness, most of the spin
labels will be suppressed or represented schematically, as in
equation~\eqref{groupint}, in the rest of the paper.

Starting from this basic setup, new models may be obtained by modifying
the partition function directly, by changing amplitudes at the level of
equation~\eqref{sf-partf}, or at an intermediate level, by modifying the
integrands in equation~\eqref{groupint}. We present the new models
following the last approach. Note that the subsequent discussion
emphasizes not the derivation of these models from first principles, but
their formulation in a unified framework suitable for computation.

\subsection{Gravity, Barrett-Crane and new models}
The Barrett-Crane (BC) model starts with the quantized $BF$ theory path
integral~\eqref{sf-partf} and imposes restrictions on the spin foam
summation in equation \eqref{sf-partf}. These restrictions heuristically
correspond to imposing the simplicity constraints on the $B$
field~\cite{BC-riem}. The restriction is twofold. First, the $\Spin(4)$
representations are restricted to balanced ones $\bj=(j,j)$, where
$j^-=j^+=j$. Second, the intertwiner summation and edge weights of
equation~\eqref{groupint} are modified such that the $i$-sums contain
only a single term corresponding to the so-called BC $4$-valent
intertwiner.

The BC model amplitudes are given in section~\ref{BC-sec}.  The
evaluation of this vertex amplitude is discussed in several
papers~\cites{CE,BCE,BCHT,KC}, where variations on the face and edge
amplitudes have also been considered.

Recently, shortcomings of the BC model have been identified by several
authors. Specifically, while this vertex amplitude correctly reproduced
the asymptotic behavior of some graviton propagator matrix elements, it
does not do so for all of them~\cites{R-prop,AR-I,AR-II}. Modified spin
foam models, referred to here as \emph{new models}, have been
subsequently proposed with the hope of overcoming these difficulties.
The model proposed by Engle, Pereira, and Rovelli
(EPR)~\cites{EPR,EPR-long} and by Livine and
Speziale~\cites{LS-coh,LS-sol} had the common motivation of identifying
its boundary state space with the space of spin network states of loop
quantum gravity. The model proposed by Freidel and Krasnov
(FK)~\cite{FK} was derived in a similar fashion, but made different
choices while imposing the simplicity constraints. As a result, the FK
model's boundary state space is different from that of the EPR one. More
recently, Conrady and Freidel have discussed in more detail the boundary
state space of the FK model~\cite{CF}.

\subsection{Model framework}
The BC, EPR, and FK models may be presented within the same framework,
following~\cite{FK}. We briefly present this framework and how each
model is realized in it.

The first step, compared to $BF$ theory, as for the BC model, is to restrict the
$\Spin(4)$ representations to balanced ones, $\bj=(j,j)$.

Consider a single strand from the double integral in
equation~\eqref{groupint}. It depicts the product of two linear
operators, corresponding to group elements $g$ and $g'$, in the
$\Spin(4)$ irrep $\bj$. One could always insert the identity operator
between $g$ and $g'$ without changing anything. On the other hand,
inserting a different linear operator in the same place will produce
different results. Keeping with the goal of identifying the
$i$-intertwiners in~\eqref{groupint} with the $SU(2)$ intertwiners in
loop quantum gravity spin networks, instead of inserting an arbitrary
linear operator, we insert only an arbitrary $SU(2)$ intertwiner. The
difference between the spin foam models described in the following
sections comes down to the choice of this insertion. An $SU(2)$
intertwiner is inserted in the following sense.

Because of the decomposition $\Spin(4)\isom SU(2)\times SU(2)$, a
balanced irrep of $\Spin(4)$ can be written as the tensor product of two
copies of an $SU(2)$ irrep, $\bj = (j,j) = j\otimes j$. Seen as a tensor
product of two $SU(2)$ reps, using Clebsch-Gordan rules, it can be
decomposed into a direct sum of $SU(2)$ representations $0$, $2$,
\ldots, $2j$.  This decomposition is not unique, since one is possible
for each diagonal $SU(2)$-subgroup of $\Spin(4)$. However, each such
choice is equivalent because of the group integrals surrounding the
operator insertion [cf.\ \eqref{groupint}]; ultimately, spin foam
amplitudes are independent of the choice.  By inserting an $SU(2)$
intertwiner, $j\otimes j\to j\otimes j$, between $g$ and $g'$, we
essentially insert a linear combination of
projections onto the $SU(2)$ irreducible invariant subspaces:
\def\sb#1{\scalebox{.85}{#1}}
\begin{equation}\label{opins}
	\sb{\idprlone}~~ = ~~\sb{\idprltwo}~~~ =
	~\sb{$\displaystyle \phantom{\sum_k \frac{\loopsn \!\!\!2k}{\thetasn
	\!\!j^*,2k}}\recprlnosum$}
	~~ \longrightarrow
	~\sb{$\displaystyle \phantom{\sum_k
	C_{jk}\frac{\loopsn}{\thetasn}}\recprl\phantom{C_{jk}}$} \!\!,
\end{equation}
where the $C_{jk}$ are weight factors parametrizing the choice of
intertwiner and the trivalent vertex corresponds to the Clebsch-Gordan
tensor. The extra coefficients are necessary, in our choice of
normalization, for the Clebsch-Gordan decomposition of the $j\otimes j$
representation.

Each strand in the above diagram carries a spin label. For compactness
of presentation, some labels are listed separated by commas, next to a
network (instead of being directly attached to the corresponding strand)
or even omitted. Some groups of spins are also represented by a
collective label like $j^*$. No ambiguity should arise, as the spin
labelling is essentially unique, given the equality relations. This
convention is used throughout the rest of the paper.

The overall spin network conventions and normalizations used in this
paper are those of~\cites{KL,CFS}. Their precise relation to
$SU(2)$-tensor contractions is presented in detail in the Appendix
of~\cite{CCK}. Note that the intermediate spin%
	\footnote{The notation of this paper differs from reference~\cite{FK},
	as Freidel and Krasnov use half-integral spins, while we use integral
	twice-spins to label $SU(2)$ irreps. In the present notation, their $j$
	becomes $j/2$, but their $k$ does coincide numerically with our $k$.
	Also, their $l$ corresponds to $i$ introduced in
	equation~\eqref{tripetal-def}.} %
is always even and so can be written as $2k$ for integer $k$.

Consider just a single Clebsch-Gordan projector insertion for each of
the strands in~\eqref{groupint}, as show in~\eqref{opins}, and
concentrate only on the part of that equation below the dotted line.
The same calculation will have to be done above the dotted line, that is
on, both sides of each tetrahedron in the triangulation.  The $\Spin(4)$
group integral can be written as an integral over two copies of $SU(2)$,
where $g = (g^-,g^+)$. The matrix elements of $g$ in a balanced
representation $\bj$ will be an outer product of matrix elements of
$g^-$ and $g^+$ in representation $j$ (circles with $-$ and $+$,
respectively):
\def\sb#1{\scalebox{.81}{#1}}
\begin{equation}\label{tripetal-def}
\sb{$\displaystyle
	\phantom{\int \d{g^-}\d{g^+}}
	\gapre~~~ = \!\!\!\!\phantom{\displaystyle \sum_{i,i^-,i^+}} \gapost~~.
$}
\end{equation}
Summations over the $i^\pm$ intertwiners again follow directly from the
property that group integration is equivalent to projection onto the
space of intertwiners between the four $j$ representations
(\emph{$j$-spins}). The extra summation over the $i$ intertwiners can be
inserted because the Clebsch-Gordan projectors map each pair of $i^\pm$
intertwiners into the subspace of intertwiners between the four $2k$
representations (\emph{$k$-spins}). These intertwiners can be
conveniently parametrized, as depicted, by an even integer $2i$
(\emph{$i$-spins})%
	\footnote{It should be noted that reference~\cite{EPR} uses
	half-integral spins, while we use integral twice-spins to label
	$SU(2)$ irreps.  However, Engle, Pereira and Rovelli's definitions for
	$i$- and $j$-spins numerically coincide with ours.}. %

The open spin networks at the bottom of the right hand side
of~\eqref{tripetal-def} join with other similar spin networks and form
left $(-)$ and right $(+)$ pent networks, which will contribute to the
corresponding vertex amplitude.  These are spin networks with topology
shown in~\eqref{fifij}, obtained by substitution of $i^-$ and $i^+$
intertwiners into the pent graph of equation~\eqref{penteq}.  The open
spin network at the top of the same expression joins with its mirror
image above the dotted line and contributes to the corresponding edge
amplitude. With the exception of the \emph{tripetal spin network}%
	\footnote{This spin network was first introduced in equation~(5)
	of~\cite{EPR} and is also referred to as a \emph{fusion coefficient}.
	The name \emph{tripetal} is suggestive of the topology of the network,
	in which the spins $i^+$, $i^-$ and $2i$ correspond to the edges of
	the three petals.} %,
located at the center of the diagram on the right
of~\eqref{tripetal-def}, all spin networks appearing so far have known
evaluations. They have come up in the evaluation of the BC vertex
amplitude and have been explicitly computed using recoupling techniques
from~\cites{KL,CFS}. The tripetal spin network will be evaluated in
section~\ref{new-eval}.

To completely define each of the three models under consideration, it
remains only to specify the dual $2$-skeleton spin foam labelling and
the weight factors $C_{jk}$ in~\eqref{opins}. The face, edge and vertex
amplitudes are then specified by the preceding construction (see
section~\ref{new-eval} for an important caveat). Most generally, in this
framework, the spin foams summed over in the partition
function~\eqref{sf-partf} assign a $j$-spin to each dual face, an
$i$-spin to each dual edge, and a $k$-spin to each dual edge-dual face
pair. However, the number of labels may be reduced in particular models.
At present, only the FK model uses all label types.

As mentioned earlier, while this presentation is convenient for
computational purposes, it hides some of the motivation from the
derivation of these models. More physical insight for each model can be
found in the original references.

\subsubsection{BC model\label{BC-sec}}
In the Barrett-Crane model, the faces of the dual $2$-complex are
labelled by $j$-spins.  The choice of intertwiner insertion weights are
$C_{jk} = 0$ ($k\ne 0$), and $C_{j0} = (-)^j(j+1)$, which is the value
of the $j$-loop spin network. Each dual edge is shared by $4$ dual
faces, while each dual vertex is shared by $10$ dual faces. The
preceding construction specifies the following dual face, edge and
vertex amplitudes:
\begin{subequations}
\begin{equation}
	A_f(F) = (j_f+1)^2, \quad A_e(F) = 1, \quad\text{and}\quad A_v(F) = \tenj,
\end{equation}
where
\begin{equation}
\label{BC-basis}
	\bclhs ~~=~~ \sum_i \frac{\loopsn \!\!\!\!i}{\thetasn
	\!\!\thetasn \!\!j^*,i}
	~\begin{smallmatrix}\bcrhs\\\bcrhs\end{smallmatrix}~.
\end{equation}
\end{subequations}
Here the spin arguments are determined by the dual faces sharing the
given cell of the $2$-skeleton. Specifically, the vertex amplitude
depends on $10$ spins, hence its name, the BC $10j$-symbol. It is
important to note that different edge and face amplitudes have been
proposed for the BC model as well~\cites{PeRo,DFKR,BCHT}.

\subsubsection{EPR model}
In the Engle-Pereira-Rovelli model, the dual faces are labelled by
$j$-spins, and dual edges are labelled by $i$-spins
[cf.~\eqref{tripetal-def}]. The weights are $C_{jj} = 1$ and $C_{jk} =
0$ for $k\ne j$. Each dual edge is shared by $4$ dual faces, while each
dual vertex is shared by $10$ dual faces and $5$ dual edges.  The
preceding construction specifies the following dual face, edge and
vertex amplitudes:
\def\sb#1{\scalebox{.97}{#1}}
\begin{subequations}
\begin{equation}
	A_f(F) = (j_f+1)^2, \quad
	A_e(F) = \frac{\loopsn\!\!\!\!2i_e}{\thetasn \thetasn \!\!2j^*,2i_e},
\end{equation}
and
\begin{equation}
	A_v(F) = \tenji,
\end{equation}
where
\begin{equation}
\label{EPR-basis}
\sb{$\displaystyle
	\bclhsi = \sum_{i^-,i^+}
	\frac{\loopsn \!\!\!\!i^-\!\!\loopsn \!\!\!\!i^+\scalebox{.75}{\tripetalepr}}
		{\thetasn\!\!\thetasn \!\!j,i^- \thetasn\!\!\thetasn \!\!j,i^+}
	\begin{smallmatrix}\bcrhsm\\\bcrhsp\end{smallmatrix}.
$}
\end{equation}
\end{subequations}
Here the spin arguments are determined by the dual faces and edges
sharing the given cell of the $2$-skeleton. Specifically, the vertex
amplitude depends on the $10$ $j$-spins from the dual faces sharing it,
as well as the $5$ $i$-spins from the incident dual edges, hence it may
be called the EPR $15j$-symbol. The same vertex amplitude was derived
in~\cite{EPR} and~\cite{LS-sol}, although the former reference was not
specific about face and edge amplitudes.

\subsubsection{FK model}
In the Freidel-Krasnov model, the dual faces are again labelled by
spins, denoted $j$, and dual edges are also labelled by intertwiners,
denoted $i$, and finally each dual edge-face pair contributes an
independent spin, denoted $k$ [cf.~\eqref{tripetal-def}]. The weight
factor is more complex than for the other two models and is given by
\begin{equation}
	C_{jk} = \frac{[(j+1)!]^2}{(j-k)!(j+k+1)!}.
\end{equation}
Each dual edge is shared by $4$ dual faces, while each dual vertex is
shared by $10$ dual faces and $5$ dual edges, as well as $20$ individual
dual edge-face pairs.  The preceding construction specifies the
following dual face, edge and vertex amplitudes:
\def\sb#1{\scalebox{.97}{#1}}
\begin{subequations}
\begin{equation}
	A_f(F) = (j_f+1)^2, ~~
	A_e(F) = \frac{\loopsn\!\!\!\!2i_e}{\thetasn\!\!\thetasn \!\!2k^*,2i_e}
		\prod_{f\supset e} C_{j_{f}k_{f}} \frac{\loopsn\!\!\!\!2k_f}{\thetasn\!\!j^*_f,2k_f},
\end{equation}
and
\begin{equation}
	A_v(F) = \tenjik,
\end{equation}
where
\begin{equation}
\label{FK-basis}
\sb{$\displaystyle
	\bclhsik = \sum_{i^-,i^+}
	\frac{\loopsn \!\!\!\!i^-\!\!\loopsn \!\!\!\!i^+\scalebox{.75}{\tripetal}}
		{\thetasn\!\!\thetasn \!\!j^*,i^- \thetasn\!\!\thetasn \!\!j^*,i^+}
	\begin{smallmatrix}\bcrhsm\\\bcrhsp\end{smallmatrix}.
$}
\end{equation}
\end{subequations}
Here the spin arguments are determined by the dual faces and edges
sharing the given cell of the $2$-skeleton. Specifically, the vertex
amplitude depends on the $10$ $j$-spins from the dual faces sharing it,
as well as the $5$ $i$-spins from the incident dual edges, and on the
$20$ $k$-spins from the dual edge-face pairs sharing it. Thus it may be
called the FK $35j$-symbol. Setting all $k$-spins, and necessarily all
$i$-spins, to $0$, this model exactly reproduces the BC spin foam
amplitudes. Also, setting all $k$-spins equal to the corresponding
$j$-spins exactly reproduces the EPR vertex amplitude $A_v(F)$. However,
in that case, the EPR edge amplitude $A_e(F)$ is reproduced with the
extra factor $\prod_{f\supset e} [(j_f+1)!]^2/(2j_f+1)!$.

Here is a brief summary of the spin labelling conventions: The BC model
assigns integer labels only to dual faces (\emph{$j$-spins}). The EPR
model also assigns integer labels to dual edges (\emph{$i$-spins}). The
FK model additionally assigns integers to each dual edge-dual face pair
(\emph{$k$-spins}). The full labelling is illustrated explicitly for a
single pent graph in figure~\ref{pent}. A detailed explanation of label
notation follows at the end of the next section.

\subsection{Observables and boundary states\label{sf-bdr}}
If the underlying triangulated manifold is closed, then corresponding
spin foams are also said to be \emph{closed}. Similarly, if the
underlying manifold has a boundary, the spin foams are said to be
\emph{open} and also have a boundary. Any open spin foam $F_O$ can be
decomposed into $F_O = F \cup F_B$, where $F_B$ labels only cells dual
to the boundary, while $F$ labels only cells dual to triangulation
simplices in the interior.  For an open spin foam $F_O$, its amplitude
may be naturally generalized to
\begin{equation}
	A(F_O) = A(F,F_B) \Psi(F_B),
\end{equation}
where the bulk amplitude $A(F,F_B)$ is the usual amplitude defined
according to~\eqref{sf-amp}, and $\Psi$ is referred to as the boundary
state, which may be fixed separately from the bulk amplitude.  The
partition function in the presence of a boundary is then written as
\begin{equation}
	Z_\Psi = \sum_{F,F_B} A(F,F_B) \Psi(F_B).
\end{equation}
Observables are functions $O(F)$ defined on spin foams. Their
expectation values, in the absence and in the presence of a boundary,
are respectively
\begin{equation}\label{bdr-oev}
	\langle O \rangle = \frac{1}{Z} \sum_F O(F) A(F)
		\quad\text{and}\quad
	\langle O \rangle_\Psi
		= \frac{1}{Z_\Psi} \sum_{F,F_B} O(F,F_B) A(F,F_B) \Psi(F_B).
\end{equation}

As an illustration, an open spin foam model with a boundary state may
arise if we split a closed spin foam model in two parts and average over
one of them. Suppose a close triangulated manifold can be decomposed
into two bulk pieces and the codimension-$1$ boundary between them. Any
closed spin foam $F_C$ can then be decomposed as $F_C = F \cup F_B \cup
F'$, where $F_B$ corresponds to the boundary, $F$ to the interior of the
piece we are interested in and $F'$ to the interior of the other piece.
The partition function may be rewritten as follows:
\begin{multline}
	Z = \sum_{F_C} A(F,F_B) A(F_B) A(F',F_B) \\
	= \sum_{F,F_B} A(F,F_B) A(F_B) \sum_{F'} A(F',F_B)
	= \sum_{F,F_B} A(F,F_B) \Psi(F_B),
\end{multline}
where $\Psi(F_B)$ has been defined by averaging over all spin foams
$F'$. This example is very similar to the separation of a large system
into a subsystem and the environment in quantum statistical mechanics.

The simplest example of a triangulation with boundary is a single
$4$-simplex, with the five tetrahedra forming its boundary. The
$2$-complex dual to the interior consists of a single dual vertex,
corresponding to the $4$-simplex itself. The dual $2$-complex of the
boundary consists of five dual edges, dual to the tetrahedra, and of ten
dual faces, dual to the triangular faces of the tetrahedra. The problems
described in sections~\ref{wavep} and~\ref{gprop}, have previously only
been considered for a single $4$-simplex. This paper restricts attention
to the same case.

\begin{figure}
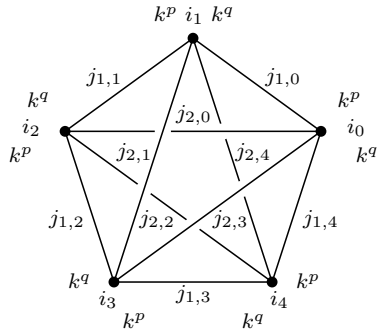

\begin{center}
	\pentijk
	\caption{The \emph{pent graph}, summarizing the indexing scheme for
	$i$-, $j$-, and $k$-spins.\label{pent}}
\end{center}
\end{figure}

The algorithms that will be described in section~\ref{num-alg} are
applicable only to a restricted class of states, \emph{factored} states.
Such a state must factor in a specific way with respect to the spins it
depends on. The various spin labels of the dual complex of the
$4$-simplex and the corresponding notation are summarized on the pent
graph of figure~\ref{pent}.
The vertices of the pent graph correspond to the five boundary
tetrahedra of the $4$-simplex, while the ten edges connecting them
correspond to its triangles. This graph is labelled by $35$ spins,
$i_e$, $j_{c,e}$, and $k^x_{c,e}$. The $e$ subscript numbers the
vertices of the pent graph; it is always taken mod~$5$. The spin
$j_{c,e}$ labels the graph edge joining vertices $e$ and $e+c$. The
superscript $x$ stands for either $p$ or $q$; $k^p_{c,e}$ labels the
vertex-edge pair $e$ and $(c,e)$, while $k^q_{c,e}$ labels the pair
$e+c$ and $(c,e)$. Again, all vertex indices are taken mod~$5$.

The class of factored states is somewhat different for each model.
However, it contains at least all of the following:
\begin{equation}\label{fac-state}
	\Psi(F_B) = \prod_{c,e} \psi_{c,e}(j_{c,e}) \prod_e \psi_e(i_e)
		\prod_{x,f,e} \psi_{x,f,e}(k^x_{f,e}),
\end{equation}
where the products range over all $i$-, $j$-, and $k$-spins. Spins not
part of a particular model may be dropped from the product.  The $\psi$s
are arbitrary functions with finite support (more on that below). For
each model, the class of factored states is enlarged, as factors of
$\Psi(F_B)$ may be allowed to depend on specific clusters of spins,
instead of only individual ones. The details will be elaborated in
section~\ref{num-alg}.

Nearly all previous work on the problems described in
sections~\ref{wavep} and~\ref{gprop} has considered only
factored boundary states. While this class of states is restrictive, its
limitations may be overcome. Note that the expectation value $\langle O
\rangle_\Psi$ in equation~\eqref{bdr-oev} is equal to the ratio of two
quantities that are both linear in the boundary state $\Psi$.  The
numerical algorithm computes this numerator and denominator separately.
Since any boundary state $\Psi$ can be approximated by linear
combinations of factored states with finite support, so can $\langle O
\rangle_\Psi$ be approximated for any boundary state $\Psi$.
However, the efficiency of such an approach is yet to be analized. The
condition of finite support for the factors $\psi(j)$ is crucial for the
sums defining $\langle O \rangle_\Psi$ to be finite.

\section{Observables and asymptotics\label{obsv-asym}}
One of the motivations for constructing new vertex amplitudes is the
recently discovered inadequacy of the BC model in reproducing
semiclassical graviton propagator behavior in the large spin limit.
Some of the propagator matrix elements show the expected behavior, while
others do not~\cites{R-prop,AR-I,AR-II}. Thus, it is important to
identify where the new vertex models differ from the BC one and whether
they have better semiclassical limits.

The comparison should ultimately be done at the level of physical
observables computed within each model. An important class of
observables, already mentioned above, are matrix elements of the
graviton propagator. They will be discussed below in
section~\ref{gprop}. Another class of observables associated with
propagation of semiclassical wave packets is addressed in
section~\ref{wavep}. Finally, a possibly less physically meaningful but
technically simpler comparison can be made at the level of vertex
amplitudes. It too can reveal important information about the behavior
of the models and is described first below.

\subsection{Comparison of asymptotics\label{bc-compar}}
One complication is the difference in the spin argument structures: the
BC vertex has 10 spin arguments, the EPR vertex has 15 spin arguments,
while the FK vertex has a total of 35 independent spin arguments. This
complication may be overcome by fixing the 10 common $j$-spin arguments
and maximizing the vertex amplitude over the remaining spins. This
effective vertex amplitude can be substituted into the partition
function~\eqref{sf-partf} where the summation is then performed over
spin foams which only assign $j$-spin labels to the dual $2$-complex.
This simplification allows the comparison of amplitudes for individual
spin foams.

It is important to note that the amplitudes in~\eqref{sf-partf} have
contributions from faces and edges as well as vertices.  The face
amplitudes are the same for all models and are easily factored out. The
edge amplitudes, on the other hand, also differ from model to model and
thus must be included in the amplitude comparison. To make the
comparison on a vertex by vertex basis, the edge amplitudes are split
between the vertices they connect as follows:
\begin{equation}\label{Aeff-def}
	A^{\mathrm{eff}}_v(j) \sim \max_{i,k} A_v(j,k,i)
		\sqrt{\prod_{e\supset v} |A_e(j,k,i)|}.
\end{equation}
For simplicity, we consider only spin inputs where each of the
$j_{c,e}$, $i_e$ and $k^{x}_{c,e}$ sets of spins have equal values,
respectively denoted by $j$, $i$, and $k$. Our assumption is that vertex
amplitudes for these spin inputs behave generically.  Small scale
numerical tests support this assumption. Otherwise, maximizing the
expression in~\eqref{Aeff-def} over a larger $i,k$-parameter space
quickly becomes impractical.

\subsection{Semiclassical wave packets\label{wavep}}
The problem presented in this section was introduced in~\cite{MPR}.
Consider a single $4$-simplex. As shown in the preceding section, it is
described by a spin foam with a single dual vertex and $i$-, $j$-, and
$k$-spins labelling cells dual to its boundary. An arbitrary functional
$\Psi(F_B)$ depending on these boundary spins, in general, corresponds
to a statistical quantum state, that is, a density matrix.

This is analogous to the single point particle, where an arbitrary
density matrix $\rho(x_f,x_i) = \langle x_f,t_f| \rho |x_i,t_i\rangle$
can be described in terms of its matrix elements between eigenstates of
the Heisenberg position operator at different times%
	\footnote{In this representation, the functional $\rho(x_f,x_i)$ is
	not necessarily symmetric, $\rho(x_i,x_f)\ne\rho(x_f,x_i)^*$.}, %
$x(t_i)|x_i,t_i\rangle = x_i|x_i,t_i\rangle$ and $x(t_f)|x_f,t_f\rangle
= x_f|x_f,t_f\rangle$.  The density matrix is pure only if it can be
factored, $\rho(x_f,x_i) = \psi(x_f,t_f)^*\psi(x_i,t_i)$, where
$\psi(x,t)$ denotes the time evolution of a given wave function.

Similarly, we can split the boundary of the $4$-simplex into two pieces%
	\footnote{Technically speaking, this decomposition is unique only in
	Lorentzian signature. In Riemannian signature, different choices of
	the decomposition should correspond to different possible Wick rotations.}, %
the initial $(-)$ and the final $(+)$. Then, for a pure boundary state,
we should be able to write
\begin{equation}\label{psi-fac}
	\Psi(F_B) = \Psi_+(F^+_B)^* \Psi_-(F^-_B),
\end{equation}
where $F^\pm_B$ respectively depend only on spins labelling the dual
complex of the corresponding piece of the boundary.

The relationship of the two boundary state factors $\Psi_\pm(F^\pm_B)$
is constrained in two ways. On the one hand (in the limit of $\hbar\to
0$), the amplitude should be peaked on those geometries that correspond
to the boundary of a classical $4$-geometry satisfying Einstein's
equations. On the other hand, $\Psi_+$ should be a time-evolved,
``future'' version of the ``past'' $\Psi_-$, which can be expressed as
\begin{equation}\label{psi-prop}
	\Psi_+(F^+_B) = \sum_{F,F^-_B} A(F,F_B) \Psi_-(F^-_B),
\end{equation}
where the summation over the boundary spin foams keeps $F^+_B$ fixed and
varies $F^-_B$.

Reference~\cite{MPR} has proposed an expression for $\Psi(F_B)$, in the
context of the EPR model, which should reproduce a flat regular
$4$-simplex. This state has gaussian dependence on individual spins and
hence is factorable in a convenient way. The problem is then to compute
$\Psi_+(F^+_B)$ both from~\eqref{psi-fac} and from~\eqref{psi-prop}, and
to compare the two. Agreement is interpreted as evidence of a correct
semiclassical limit for the EPR model.

A concrete expression for the proposed $\Psi(F_B)$ is
\begin{align}
\label{epr-bstate}
	\Psi(F_B) &= N \prod_{c,e}\psi_{c,e}(j_{c,e}) \prod_e\psi_e(i_e,\{j_{c,e}\}),
		\quad\text{with}\\
\label{bc-gauss}
	\psi_{c,e}(j_{c,e}) &= e^{-\frac{1}{\tau}(j_{c,e}-j_0)^2
			+ i\Theta j_{c,e}}, \\
\label{epr-gauss}
	\psi_e(i_e,\{j_{c,e}\}) &=
		\sqrt{\frac{2i_e+1}
			{\theta(2i_e,2j_{1,e},2j_{2,e})\theta(2i_e,2j_{1,e-1},2j_{2,e-2})}}
		e^{-\frac{3}{4j_0}(i-i_0)^2 + i\frac{\pi}{2} i_e}
	,
\end{align}
where $N$ is a normalization factor, $j_0$ determines the size of the
regular $4$-simplex and $\cos\Theta = -1/4$. The parameter $\tau$
controls the size of quantum fluctuations about the classical values of
$j$.

The wave packet propagation geometry given in~\cite{MPR} fixes $\tau=0$
in the state~\eqref{bc-gauss}, freezing all $j$-spins to the background
value $j_0$. Effectively, only the dependence of $\Psi(F_B)$ on the
$i$-spins was considered.  A single vertex of the $4$-simplex is
labelled as ``past'', while the remaining four as ``future''. The four
``future'' vertices form a tetrahedron, whose dual is labelled by an
$i$-spin. This labelled dual edge constitutes $F_B^+$, while the
remaining four dual edges labelled by $i$-spins constitute $F_B^-$. This
propagation geometry will be referred to as \emph{EPR 4-1 propagation}.

An immediate generalization, feasible with the algorithm described in
section~\ref{num-alg}, is to relax the $\tau=0$ limitation. The choice
of $\tau$ should be consistent with the parameters used in the graviton
propagator calculations. Thus, following%
	\footnote{It should be noted that reference~\cite{CLS} uses
	half-integral spins, while we use integral twice-spins to label
	$SU(2)$ irreps. A $j$ label from Christensen, Livine and Speziale
	corresponds numerically to $j/2$ in current notation.} %
\cite{CLS}, we let the wave packet width depend on the background spin,
\begin{equation}\label{tau-def}
	\tau = 4j_0/\alpha,
\end{equation}
with $\alpha$ is a positive parameter.

It is important to note that the boundary state proposed above, in
equation~\eqref{epr-bstate}, is not the best possible candidate to
generalize the calculations of~\cite{MPR}. Its major advantage is that
it belongs to the class of factored states.  The major disadvantage of
factored states is that they neglect possible spin-spin correlations.
In fact, a more realistic proposal for a boundary state where both $i$-
and $j$-spins are allowed to vary and which includes such correlations
was given by Rovelli and Speziale~\cite{RS}. Unfortunately, Rovelli and
Speziale's boundary state does not factor nicely and would require more
sophisticated techniques to be used efficiently. As such, the boundary
state proposed above should be seen more as a test of the algorithms
presented in section~\ref{num-alg} and an attempt to explore the
qualitative effects of introducing $j$-dependent (and below
$k$-dependent) boundary states.

Now, a uniform methodology should be constructed for each of the three
models. As only $j$-spins are common among the models, we propose the
following wave packet propagation geometry. One possibility is to
propagate wave packets from nine of the $j$-spins to the remaining one.
This configuration corresponds to fixing a single triangle (defined by
three vertices of a $4$-simplex) in the ``future'', while relegating the
other nine triangles (containing at least one of the two remaining
$4$-simplex vertices) to the ``past''. Thus, the single $j$-labelled
face dual to the ``future'' triangle will constitute $F_B^+$, while the
rest of the boundary spin foam will constitute $F_B^-$, including all
$i$- and $k$-spins, if any. This propagation geometry will be referred
to as \emph{9-1 propagation}.

Another alternative is to assign a vertex of the $4$-simplex to the
``future'', together with the six triangles sharing sharing it. The
remaining four triangles are relegated to the ``past''. Thus, $F_B^+$
consists of the six $j$-labelled faces dual to the ``future'' triangles,
with the rest of the boundary spin foam constituting $F_B^-$. This
propagation geometry will be referred to as \emph{4-6 propagation}.
There are numerous other possibilities. However, the two described above
are sufficient to illustrate an application of the numerical algorithms
and to show the qualitative behavior to be expected from propagated wave
packets.

The boundary state~\eqref{epr-bstate} is valid only for the EPR model.
For the BC model, we simply drop the $\psi_e$ factors:
\begin{equation}\label{bc-bstate}
	\Psi(F_B) = N \prod_{c,e}\psi_{c,e}(j_{c,e}).
\end{equation}
And for the
FK model we must add extra $\psi^x_{c,e}$ factors for each $k$-spin:
\begin{equation}\label{fk-bstate}
	\Psi(F_B) = N \prod_{c,e}\psi_{c,e}(j_{c,e}) \prod_e\psi_e(i_e,\{j_{c,e}\})
		\prod_{x,c,e} \psi^x_{c,e}(k^x_{c,e},j_{c,e}).
\end{equation}
Because the $k$-spins are closely geometrically associated with
$j$-spins, we use the same gaussian state parameters:
\begin{align}\label{fk-gauss}
	\psi^x_{c,e}(k^x_{c,e},j_{c,e}) &=
	\sqrt{\frac{2k^x_{c,e}+1}{\theta(j_{c,e},j_{c,e},2k^x_{c,e})}
		C_{j_{c,e} k^x_{c,e}}}
	e^{-\frac{\alpha}{4j_0}(k^x_{c,e}-j_0)^2 + i\Theta k^x_{c,e}}, \\
	C_{jk} &= \frac{(j+1)!}{(j-k)!} \frac{(j+1)!}{(j+k+1)!}.
\end{align}
The square root factor includes the FK model edge normalization, as
does~\eqref{epr-gauss} for the EPR model.

\subsection{Graviton propagator\label{gprop}}
The graviton propagator is well defined in the perturbative quantization
of gravity. It is computed as the $2$-point function
$G_{\mu\nu\rho\sigma}(x,y) = \langle0| h_{\mu\nu}(x) h_{\rho\sigma}(y)
|0\rangle$, where $|0\rangle$ is the Minkowski vacuum, and
$h_{\mu\nu}(x)$ is the metric perturbation. General relativity requires
that, in harmonic gauge~\cite{Wald}, the decay rate of the $2$-point
function, for large separation between points $x$ and $y$, is the same
as for the Newtonian force of gravitational attraction: inverse distance
squared.  The framework for computing the equivalent of the graviton
propagator in the spin foam formalism was elaborated
in~\cites{MR,R-prop,BMRS,LS-grint}. The quantum area spectrum is
$A=\ell_P^2(j+1)$, with $j$ a dual face spin foam label and $\ell_P$ the
Plank length. Dimensional arguments then give the expected decay of the
propagator as $O(1/j)$, with $j$ being the typical size for the chosen
spin foam boundary state.

The expected asymptotic behavior of the graviton propagator has been
checked for the BC model both analytically and
numerically~\cites{R-prop,CLS,LS-grint}.  Unfortunately, the expected
behavior was only reproduced for certain tensor components of
$G_{\mu\nu\rho\sigma}(x,y)$, but not for others~\cites{AR-I,AR-II}. This
negative result has prompted the introduction of EPR and FK spin foam
models as alternatives to the BC model. The challenge is to compute the
graviton propagator for the new models and check that it has the
expected asymptotic behavior.

Following~\cite{CLS}, we show the computational set up for the BC model
and then extend it to other models. Consider again a single $4$-simplex
with boundary and the corresponding spin foam. We associate the area $A
= \ell_P^2(j+1)$ to each triangle, depending on the $j$-spin labelling
its dual. The goal is to compute the correlation between observables
depending on the triangle areas [cf.~\eqref{bdr-oev}]:
\begin{equation}\label{gprop-def}
	W_{ce,c'e'} = \frac{1}{Z_\Psi} \sum_{F,F_B} A(F,F_B) h_{ce}(F_B)
		h_{c'e'}(F_B) \Psi(F_B),
\end{equation}
where $ce$ and $c'e'$ index the specific $j_{c,e}$ and $j_{c',e'}$ spins
taking part in the correlation. Again following~\cite{CLS}, the
boundary state%
	\footnote{We incorporate the ``measure'' discussed in~\cite{CLS} into the
	boundary state and pick the trivial case $k=0$.} %
is a semiclassical gaussian state peaked around a flat $4$-simplex,
whose scale is set by $j_0$:
\begin{equation}
	\Psi(F_B) = \prod_{c,e} e^{-\frac{\alpha}{4j_0}(j_{c,e}-j_0)^2
		+ i\Theta j_{c,e}},
\end{equation}
where $\cos\Theta = -1/4$, and $j_0$ sets the scale for the background
geometry, as in the previous section. The observables measure the
fluctuation of areas squared:
\begin{equation}
	h_{ce}(F_B) = \frac{(j_{c,e}+1)^2-(j_0+1)^2}{(j_0+1)^2}.
\end{equation}
Note that the product $\Psi'(F_B) = h_{ce}(F_B) \Psi(F_B)$ has exactly
the same factorizability properties as $\Psi(F_B)$. This property allows
both the numerator and denominator in~\eqref{gprop-def} to be computed
on the same footing.

Again, an important task here is the generalization of this calculation
to the EPR and FK models. This generalization essentially requires the
specification of a boundary state that describes a semiclassical state
peaked around the flat regular $4$-simplex. Since this is the same
requirement used in picking out the boundary states in the section on
wave packet propagation, simply choose the same ones.  That is, the BC,
EPR and FK boundary states are specified, respectively, by
equations~\eqref{bc-bstate}, \eqref{epr-bstate}, and \eqref{fk-bstate}.

\section{Spin network evaluations\label{new-eval}}
The second group integration identity in~\eqref{groupint} requires a
choice of basis in the space of intertwiners between four $\Spin(4)$
representations. This choice is arbitrary; however, some choices are
more convenient than others. For example, the normalization factor
in~\eqref{groupint} is simplest when the $i'$-basis is the same as the
$i$-basis. On the other hand, the choice of intertwiner basis in the EPR
model, equation~\eqref{EPR-basis}, is made such that when the
intertwiner networks are substituted into the vertex amplitude pent
graph, equation~\eqref{penteq}, the amplitude is resolved as a sum over
$15j$-symbols with the following topology:
\begin{equation}\label{fifij}
	\scalebox{.91}{\fifij} ~~~.
\end{equation}
A similar choice is made for the BC and FK models,
equations~\eqref{BC-basis} and~\eqref{FK-basis}. This topology is
required for the numerical algorithm described in section~\ref{num-alg}.

Unfortunately, the requirements of simple edge normalization factors and
the above topology requirement for each vertex amplitude are not always
compatible. For example, they are not compatible for the minimal
triangulation of the $4$-sphere.  With the current formulation of the
vertex amplitude evaluation algorithm, preference must be given to the
topology requirement.  A similar issue, referred to as ``edge
splitting,'' was encountered for spin foams on a cubic lattice
in~\cite{CCK}.

Throughout this paper, we have assumed that the topology and simplicity
of dual edge normalization requirements can be simultaneously satisfied.
This assumption is justified in the case of a single $4$-simplex, and
other simple arrangements of a small number of $4$-simplices. If this
assumption is not justified, then the dual edge amplitudes given in the
previous section will have to be modified, with the important exception
of the BC model. The edge normalization requirement for the BC model is
trivial.

\subsection{Tripetal network evaluation\label{trip-net}}
The tripetal spin network, defined in equation~\eqref{tripetal-def}, is
evaluated as follows. It is first written as a contraction of two
trivalent networks, along the strands labelled $i^-$, $i^+$ and $2i$.
\def\sb#1{\scalebox{.9}{#1}}
\begin{equation}\label{trip-kling}
	~~~\sb{\tripetal} ~~~ = ~~~ \sb{\triplong} ~~~.
\end{equation}
Each of these trivalent networks must be proportional to the unique
$SU(2)$ $3$-valent intertwiner. The proportionality constant is computed
explicitly through recoupling:
\begin{gather}
	\scalebox{.8}{\klingon}
	= \sum_n \frac{(-)^{\half(j+j'-n)}\Delta_{n}}
		{\theta(n,j,j')}
		~\scalebox{.8}{\klingoni} \\
	= \sum_n \frac{(-)^{\half(j+j'-n)}\Delta_{n}}
		{\theta(n,j,j')}
		\frac{\Tet{j & j & j' \\ i^- & n & 2k }}{\theta(i^-,n,2k )}
		\frac{\Tet{j' & j' & j \\ i^+ & n & 2k'}}{\theta(i^+,n,2k')}
		~\scalebox{.8}{\klingonii} \\
	= \sum_n \frac{(-)^{\half(j+j'-n)}\Delta_{n}}
		{\theta(n,j,j')}
		\frac{\Tet{j & j & j' \\ i^- & n & 2k }}{\theta(i^-,n,2k )}
		\frac{\Tet{j' & j' & j \\ i^+ & n & 2k'}}{\theta(i^+,n,2k')}
		\frac{\Tet{2k' & 2k & n \\ i^- & i^+ & 2i}}{\theta(i^-,i^+,2i)}
		~\scalebox{.8}{\klingoniii} ~~.
\end{gather}
The first equality recouples the crossing strands through the auxiliary
spin $n$. Other steps correspond to collapsing triangles to $3$-valent
vertices. The square brackets denote the evaluation of a tetrahedral
(\emph{tet}) spin network, while $\theta$ and $\Delta$ denote the
evaluations of the theta-like and loop spin networks seen
in~\eqref{opins} and elsewhere.  For full details, see~\cite{KL} and the
Appendix of~\cite{KC}.

After this simplification, the tripetal network is proportional to the
theta network, where we write the left copy of the above coefficient as
$P_{i^-i^+}$ and the right copy as $Q_{i^-i^+}$:
\begin{equation}
	\tripetal ~~~ = ~ P_{i^-i^+} ~~ \thetatw ~~ Q_{i^-i^+}.
\end{equation}
The network on the right hand side of the above equation is equal to the
theta network up to sign, which is specified in~\eqref{R-def}.  Both
$P_{i^-i^+}$ and $Q_{i^-i^+}$ depend on many spins.  The displayed
indices are those that will be important in section~\ref{num-alg}.

To obtain the final formulas for each vertex $e$ of the pent graph, we
make appropriate substitutions into the above expression, from each half
of the tripetal network.  We replace $i$ and $i^\pm$ by $i_e$ and
$i^\pm_e$ respectively.  In the $P$ coefficient, the spin $k$ becomes
$k^p_{1,e}$, while $k'$ becomes $k^p_{2,e}$. At the same time, the spin
$j$ becomes $j_{1,e}$, while $j'$ becomes $j_{2,e}$. Similarly, in $Q$,
the spin $k$ becomes $k^q_{2,e-2}$, $k'$ becomes $k^q_{1,e-1}$, $j$
becomes $j_{2,e-2}$ and $j'$ becomes $j_{1,e-1}$. The final formula for
the tripetal network is
\begin{subequations}\label{trip-eval}
\begin{equation}
	\label{R-def}
	R^e_{i^-_e i^+_e}
		= (-)^{\half(i_e^{-}+i_e^{+}-2i_e)}
			\frac{P^e_{i_e^{-}i_e^{+}}Q^e_{i_e^{-}i_e^{+}}}{\theta(i_e^-,i_e^+,2i_e)},
\end{equation}
\begin{multline}
	\label{P-def}
	P^e_{i_e^- i_e^+} = 
		\sum_{n_p} (-)^{\half(j_{1,e}+j_{2,e}-n_p)} \Delta_{n_p} \\
		\frac{\Tet{2k^p_{2,e}&2k^p_{1,e}&n_p\\i_e^-&i_e^+&2i_e}}
			{\theta(n_p,j_{1,e},j_{2,e})}
		\frac{\Tet{j_{1,e}&j_{1,e}&j_{2,e}\\i_e^-&n_p&2k^p_{1,e}}}
			{\theta(i_e^-,n_p,2k^p_{1,e})}
		\frac{\Tet{j_{2,e}&j_{2,e}&j_{1,e}\\i_e^+&n_p&2k^p_{2,e}}}
			{\theta(i_e^+,n_p,2k^p_{2,e})}
		,
\end{multline}
\begin{multline}
	\label{Q-def}
	Q^e_{i_e^-i_e^+} =
		\sum_{n_q} (-)^{\half(j_{2,e-2}+j_{1,e-1}-n_q)} \Delta_{n_q} \\
		\frac{\Tet{2k^q_{1,e-1}&2k^q_{2,e-2}&n_q\\i_e^-&i_e^+&2i_e}}
			{\theta(n_q,j_{2,e-2},j_{1,e-1})}
		\frac{\Tet{j_{2,e-2}&j_{2,e-2}&j_{1,e-1}\\i_e^-&n_q&2k^q_{2,e-2}}}
			{\theta(i_e^-,n_q,2k^q_{2,e-2})}
		\frac{\Tet{j_{1,e-1}&j_{1,e-1}&j_{2,e-2}\\i_e^+&n_q&2k^q_{1,e-1}}}
			{\theta(i_e^+,n_q,2k^q_{1,e-1})}
		.
\end{multline}
\end{subequations}
Notation for the $j$- and $k$-spins is explained in the next section.

The same spin network has also been evaluated independently by
Alesci~\textit{et al}~\cite{ABMP-anl}, however their notation differs
from ours in normalization.

\section{Numerical algorithms\label{num-alg}}
The dual face and edge amplitudes, $A_f$ and $A_e$, are trivial to
compute. The difficulty lies in evaluating the dual vertex amplitude
$A_v$, which is where we will concentrate.  All algorithms described in
this section are extensions of the original CE algorithm for the BC
model~\cite{CE} and all fall into the same product-trace pattern:
\begin{equation}\label{prodtrace}
	A_v(\{j_{c,e},i_e,k^x_{c,e}\})
		= (-)^S \sum_{m^-,m^+} \phi \tr[M_4 M_3 M_2 M_1 M_0],
\end{equation}
where $(-)^S$ is an overall sign factor, $\phi$ depends only on $m^\pm$
and $M_e$ are matrices of compatible dimensions, collectively depending
on all the spins. Each of these elements may be specified separately in
any incarnation of this algorithm. In all cases presented below, we have
\begin{equation}\label{phi-def}
	\phi = (-)^{\half(m^- - m^+)} (m^- + 1)(m^+ + 1).
\end{equation}
However, the $M_e$ matrices will be redefined for each variation of the
algorithm. The notation for various boundary spins is summarized with
the pent graph in figure~\ref{pent}.

The reason for the similarity is that all the vertex amplitudes
discussed in section~\ref{bf-th} are structurally similar to the vertex
amplitude~\eqref{penteq} for discrete $BF$ theory, in which the
intertwiners can be resolved to produce $SU(2)$ spin networks of the
topology shown at the beginning of section~\ref{new-eval}. The topology of the
pent graph, as described in~\cite{CE}, is what gives rise to the product
trace structure~\eqref{prodtrace}. The differences between the $M_e$
matrices for each model capture the differences between the $\Spin(4)$
intertwiners used to define the corresponding vertex amplitudes in
equations~\eqref{BC-basis}, \eqref{EPR-basis}, and~\eqref{FK-basis}.
When contracting with a boundary state, the structure of the $M_e$
matrices also captures the extra summations with respect to the boundary
spins.

\subsection{Run time complexity}
The discussion below assumes familiarity with run time complexity
estimates of standard numerical linear algebra operations.

The run time complexity of a generalized CE algorithm may be estimated
as follows. Suppose that the spin arguments to $A_v$
in~\eqref{prodtrace} are of average magnitude $j$. Then, generally, the
dimensions of the matrices $M_e$ scale as a power of $j$; say, each
matrix is $O(j^d)\times O(j^d)$, for some integer $d$.  The run time
will be dominated by filling the $M_e$ matrices and by the product-trace
operation. This exponent $d$ must be determined separately, depending on
the model and on the inclusion of a boundary state.

The product-trace may be implemented as follows: each of the $O(j^d)$
standard basis vectors is subjected to matrix-vector multiplies by the
$M_e$ and appropriate elements of the result vectors are accumulated
into the trace. If the $M_e$ are dense, then the cost of a matrix-vector
multiply is $O(j^{2d})$.  However, we shall see below that this
complexity may be reduced by decomposing each $M_e$ into sparse factors.
Hence, we will parametrize the matrix-vector multiply complexity as
$O(j^D)$, with $D$ no greater than $2d$, and the product-trace
complexity as $O(j^{d+D})$.

If each matrix element is computed in at most $O(j^f)$ time, the upper
bound on the time needed to fill an $O(j^d)\times O(j^d)$ matrix $M_e$
is $O(j^{2d+f})$. Sparse factorization improves this estimate as well,
which we will parametrize as $O(j^{F+f})$, where $F$ also does not
exceed $2d$. In all cases we have examined, $d+D > F+f$, which implies
that the product-trace operation dominates matrix filling in run time
for large spins. More detailed discussions of possible optimizations for
matrix filling can be found in~\cite{KC}. Below, we will give the best
known value of $f$ for each algorithm. In particular, if a matrix
element depends on a tet, the standard Wigner-Racah formula requires
$O(j)$ operations to compute it ($f=1$). On the other hand, using
recurrence relations~\cite{SGrec} or hashing techniques the number of
operations can be reduced to $O(1)$ ($f=0$).

Finally, the outer $m^\pm$ sums in~\eqref{prodtrace} also span ranges of
size $O(j)$. Therefore, the run time complexity of a generalized CE
algorithm may be expressed as $O(j^{2+d+D})$. 

The matrix elements of the $M_e$ (computed for each case in the
following sections), will contain spin network evaluations that require
certain inequality and parity constraints on their arguments. Solving
these constraints yields precise matrix dimensions and bounds for any
intermediate summations. The details are described at the end of
section~\ref{fix-new-alg}. It can be shown that all matrix dimensions
as well as intermediate summation bounds are finite. However, for that
to be true in the presence of a boundary state, it is crucial that each
factor of the boundary state~\eqref{fac-state} has finite support.

\subsection{Algorithms for fixed boundary spins}
\subsubsection{BC vertex\label{fix-bc-alg}}
For the BC model, the vertex amplitude is only a function of the
$j$-spins. As a slight abuse of notation, we will use the symbols $i_e$
as indices (also referred to as spins, and directly analogous to the
$i_e^\pm$ indices introduced for the other models) of the $M_e$
matrices:
\begin{equation}\label{bc-m}
	(M_e)^{i_{e+1}}_{i_e} =
	\frac{(i_e+1) \Tet{ i_e & j_{2,e} & m^- \cr i_{e+1} & j_{2,e-1} & j_{1,e}}
			\Tet{ i_e & j_{2,e} & m^+ \cr i_{e+1} & j_{2,e-1} & j_{1,e}}}
		{\theta(j_{2,e-1},i_{e+1},j_{1,e})\,\theta(j_{2,e},i_{e},j_{1,e})\,
			\theta(j_{2,e},i_{e+1},m^-)\,\theta(j_{2,e},i_{e+1},m^+)} .
\end{equation}
The sign factor from~\eqref{prodtrace} is given by $S = \sum_{c,e}
j_{c,e}$. The ranges of the $i_e$ and $m$ spins are specified by
triangle inequalities and parity constraints satisfied by various spins.
For a detailed derivation and for notation, see the original
reference~\cite{CE}, and also~\cite{KC} and its Appendix%
	\footnote{Note however, that these references use half-integral spins,
	while the present paper uses integer twice-spins.}. %

The structure of the $M_e$ matrices will become increasingly important
and will grow in sophistication in the algorithms presented below.
Hence, it is convenient to introduce a graphical notation to represent
this structure. In this simplest case we have:
\begin{equation}
	M_e = \phantom{i_{a+1}} \bcM \phantom{i_a} .
\end{equation}
Each strand represents an index. The incoming and outgoing strands
correspond to the $i_e$ and $i_{e+1}$ indices of $M_e$ and are labelled
as such. The product-trace operation in~\eqref{prodtrace} is effected by
concatenating appropriately labelled strands. Further features of this
graphical notation will be elaborated as they are introduced.

Each matrix $M_e$ is dense and of size $O(j)\times O(j)$. According to
the discussion at the beginning of this section, we have $d=1$, $D=2$,
$f=0$, $F=2$, and $2+d+D=5$. Hence, we recover the well known $O(j^5)$
run time complexity of the original CE algorithm.

\subsubsection{New vertices\label{fix-new-alg}}
For the EPR model, the amplitude is a function of both $i$- and
$j$-spins, while the FK model is also a function of $k$-spins. Here, we
give the explicit FK formula, with the EPR version obtained by setting
$k^x_{c,e} = j_{c,e}$. The matrix elements of $M_e$ are
\begin{equation}\label{new-m}
	(M_e)^{i^-_{e+1} i^+_{e+1}}_{i^-_e i^+_e}
	= Q^{e+1}_{i_{e+1}^{-}i_{e+1}^{+}}
	(T^e_-)^{i^-_{e+1}}_{i^-_e}(T^e_+)^{i^+_{e+1}}_{i^+_e}
	\left[
	P^e_{i_e^{-}i_e^{+}} N^e_{i^-_e} N^e_{i^+_e} 
	\frac{(-)^{\half(i_e^{-}+i_e^{+}-2i_e)}}{\theta(i_e^-,i_e^+,2i_e)}
	\right]
\end{equation}
where
\begin{align}
\label{new-T}
	(T^e_\pm)^{i^\pm_{e+1}}_{i^\pm_e} &=
		\frac{\Tet{i^\pm_e&j_{1,e}&j_{2,e-1}\\i^\pm_{e+1}&m^\pm&j_{2,e}}}
			{\theta(i^\pm_{e+1},m^\pm,j_{2,e})}, \\
\label{new-N}
	N_{i^\pm_e} &= \frac{\Delta_{i^\pm_e}}
		{\theta(i^\pm_e,j_{1,e},j_{2,e}) \theta(i^\pm_e,j_{2,e-2},j_{1,e-1})},
\end{align}
while $P$ and $Q$ are given by equations~\eqref{P-def}
and~\eqref{Q-def}. A detailed discussion of the bounds on the various
spins follows below. The $T_\pm$ and $N$ terms are inherited from the CE
algorithm; $N T_- T_+$ is a particular factorization of the right hand
side of~\eqref{bc-m}. The new factors of $P$ and $Q$ come from the
evaluation of the tripetal networks in equations~\eqref{EPR-basis}
and~\eqref{FK-basis}.

The matrix elements of $M_e$ are indexed by the pairs $(i^-_e,i^+_e)$
and $(i^-_{e+1},i^+_{e+1})$. In graphical notation, $M_e$ has the
following structured factorization.
\begin{equation}\label{new-M}
	M_e = \phantom{i_{e+1}} \newM \phantom{i_e} ,
\end{equation}
where $\bar{P}^e$ stands for the entire bracketed term in~\eqref{new-m}.
Unmarked vertices in the above diagram essentially correspond to
Kronecker deltas. The notation is saying that both $P^e$ and $Q^{e+1}$
are diagonal matrices acting on the space of vectors indexed by
$(i^-_e,i^+_e)$ or $(i^-_{e+1},i^+_{e+1})$. On the other hand, the
$T^e_\pm$ matrices are block diagonal, acting separately on the $-$ and
$+$ indices. Note that the graphical notation used in~\eqref{new-M} is
not directly related to that of sections~\ref{bf-th}
and~\eqref{new-eval}. The similarity between the them reflects the fact
that one can introduce a graphical notation to represent any product of
tensors (functions of multiple arguments) contracted over some indices
(summed or integrated over some, possibly repeated, arguments).

To implement the above algorithm, it is important to compute
the precise range of the $m^\pm$ summations, the size of each $M_e$
matrix, that is, the allowed ranges of the $i_e^\pm$ spins, and the ranges
of the $n_{p,q}$ summations in the definitions of $P$ and $Q$
in~\eqref{trip-eval}.  Whenever the arguments of either the theta or
tetrahedral spin networks fail to satisfy certain conditions, these
networks evaluate to zero. Therefore, the $m^\pm$, $i_e^\pm$, and
$n_{p,q}$ ranges are taken to be the largest such that all necessary
conditions are satisfied. These conditions are
\begin{align}
	\theta(a,b,c):& ~~ \tri(a,b,c), \\
		\quad\text{and}\quad
	\Tet{a & b & e \\ c & d & f}:& ~~
		\tri(c,d,f),~\tri(a,b,f),~\tri(a,d,e),~\tri(c,b,e),
\end{align}
where the abbreviation stands for the triangle inequality and parity
constraint
\begin{equation}\label{tri-cond}
	\tri(a,b,c): ~~ a \le b+c, b \le c+a, c \le a+b, ~\text{and}~
		a+b+c = 0 \pmod{2}.
\end{equation}
It can be shown that these conditions, once collected from
equations~\eqref{trip-eval}, \eqref{new-T} and \eqref{new-N}, are
sufficient to make all summations involved in the algorithm finite.
The linearity of the triangle inequalities also implies that the upper
bound on all sums grows linearly with the magnitude of the input $i$-,
$j$-, and $k$-spins. However, it is important for an efficient
implementation to obtain the tightest possible bounds on each of the
summation indices.

The dimension of each $M_e$ is $O(j^2)\times O(j^2)$, implying $d=2$.
However, each $M_e$ decomposes into sparse (diagonal or block diagonal)
factors. The filling complexity parameters for the largest of these
factors, $\bar{P}$ and $Q$, are $f=1$, $F=2$, and $F+f=3$.  Also, the
cost of a matrix-vector multiply is parametrized by $D=2+1=3$, giving
$2+d+D=7$. Therefore, the run time complexity of evaluating an EPR or FK
vertex amplitude is $O(j^7)$. This estimate compares favorably to
simply treating $M_e$ as a dense $O(j^2)\times O(j^2)$ matrix, which
would imply an overall $O(j^8)$ run time complexity.

\subsection{Algorithms for boundary states}
Contracting a boundary state, as described in section~\ref{sf-bdr}, with
the vertex amplitude~\eqref{prodtrace} gives the following partition
function
\begin{equation}\label{bdr-contract}
	Z_\Psi = \sum_{\{j_{c,e},i_e,k^x_{c,e}\}}
		A_v(\{j_{c,e},i_e,k^x_{c,e}\}) \Psi(\{j_{c,e},i_e,k^x_{c,e}\}).
\end{equation}
A naive approach to the problem of computing $Z_\Psi$ would wrap an
algorithm to compute $A_v$ (as described in the previous section) in as
many outer sums as there are spins in $\{j_{c,e},i_e,k^x_{c,e}\}$.
Namely, for the BC model, this would produce a calculation of run time
complexity $O(j^{5+10})=O(j^{15})$, with $10$ outer spin sums. The EPR
model would yield $O(j^{7+15})=O(j^{22})$, with $15$ outer spin sums,
and the FK model $O(j^{7+35})=O(j^{42})$, with $35$ outer spin sums.
Clearly, with the naive approach, these problems become intractable.
Fortunately, when dealing with a factored state (as defined tentatively
in section~\ref{sf-bdr} and more precisely in the following sections),
these summations may be absorbed into a redefinition of the $M_e$
matrices of sections~\ref{fix-bc-alg} and~\ref{fix-new-alg}, producing
again a generalized CE algorithm:
\begin{equation}\label{CE-bdr}
	Z_\Psi = \sum_{m^-,m^+} \phi \tr[M_4 M_3 M_2 M_1 M_0] ,
\end{equation}
where $\phi$ is still defined by equation~\eqref{phi-def} and the sign
factor is necessarily absorbed into the $M_e$.  This approach is
described in the next two sections.

It is important to note that the allowed spin summation ranges, which
determine the dimensions of the $M_e$ matrices, may be strongly impacted
by the presence of a finitely supported boundary state. It is convenient
for our purposes to keep the assumption that, even in the presence of a
boundary state, the summation range for each spin is still of order
$O(j)$, an assumption justified for the boundary states proposed in
sections~\ref{wavep} and~\ref{gprop} (or finitely supported
approximations to them). The run time complexity will be analyzed only
for this case. However, the same analysis can be easily performed in
other cases, where some of the spin summation ranges are significantly
different.

\subsubsection{BC vertex with boundary states\label{bc-bdr}}
For the BC model, consider a factored boundary state of the form
\begin{equation}\label{bc-psi}
	\Psi(\{j_{c,e}\}) = \prod_{c,e} \psi_{c,e}(j_{c,e}).
\end{equation}
The dependence of the matrices given in~\eqref{bc-m} on $\{j_{c,e}\}$
allows us to obtain the form~\eqref{CE-bdr} with the following
redefinition:
\begin{multline}\label{bcb-m}
	(M_e)^{j_{2,e} i_{e+1}}_{j_{2,e-1} i_e} =
	\frac{(i_e+1)\, \psi_{2,e-1}(j_{2,e-1})}
		{\theta(j_{2,e},i_{e+1},m^-)\,\theta(j_{2,e},i_{e+1},m^+)} \\
	\sum_{j_{1,e}} \psi_{1,e}(j_{1,e})
	\frac{\Tet{ i_e & j_{2,e} & m^- \cr i_{e+1} & j_{2,e-1} & j_{1,e}}
			\Tet{ i_e & j_{2,e} & m^+ \cr i_{e+1} & j_{2,e-1} & j_{1,e}}}
		{\theta(j_{2,e-1},i_{e+1},j_{1,e})\,\theta(j_{2,e},i_{e},j_{1,e})}
	.
\end{multline}
Graphically, we represent the above equation as
\begin{equation}
	M_e = \phantom{i_{e+1}} \bcbM \phantom{j_{2,e-1}} ,
\end{equation}
where $M^{\mathrm{orig}}_e$ corresponds to the right hand side of
equation~\eqref{bc-m} and $\psi$ refer to the appropriate factors of the
boundary state~\eqref{bc-psi}. The tadpole $\psi$ shows an internal
summation over $j_{1,e}$ necessary to form the matrix elements of $M_e$.
It is shown here to highlight the location of the extra summation
insertion and the possible relation of $\psi_{1,e}$ to other spins.
Note that, without any modification to the evaluation algorithm, we can
generalize the notion of factored boundary states to include factors of
the form $\psi(j_{1,e}, j_{2,e}, j_{2,e-1})$.

Notice that in this case $M_e$ is dense and of size $O(j^2)\times
O(j^2)$. Hence, the algorithm's runtime complexity is $O(j^8)$, as
$d=2$, $D=2+2$, and $2+d+D=8$, while the filling parameters are $f=1$
and $F=4$. Interestingly enough, the tets satisfy an identity which
allows us to decompose $M_e$ into sparse factors speeding up both the
product-trace and matrix filling, thus reducing the run time complexity
to $O(j^7)$. This identity is known as
the Biedenharn-Elliot identity~\cites{KL,CFS}:
\begin{equation}\label{BE}
	\frac{\Tet{A&B&C\\a&b&c} \Tet{A'&B'&C'\\a&b&c}}{\theta(a,b,c)}
	= \sum_{s} \Delta_s
	\frac{\Tet{s&C'&B'\\a&B&C}}{\theta(s,A,A')}
	\frac{\Tet{s&A'&C'\\b&C&A}}{\theta(s,B,B')}
	\frac{\Tet{s&B'&A'\\c&A&B}}{\theta(s,C,C')}
	.
\end{equation}
The product of two tets in equation~\eqref{bcb-m} can be rewritten using
this identity as
\begin{multline}
	\frac{\Tet{ i_e & j_{2,e} & m^- \cr i_{e+1} & j_{2,e-1} & j_{1,e}}
			\Tet{ i_e & j_{2,e} & m^+ \cr i_{e+1} & j_{2,e-1} & j_{1,e}}}
		{\theta(j_{2,e-1},i_{e+1},j_{1,e})} \\
	= \sum_{s_e} \Delta_{s_e}
		\frac{\Tet{s_e&m^+&j_{2,e}\\i_{e+1}&j_{2,e}&m^-}}{\theta(s_e,i_e,i_e)}
		\frac{\Tet{s_e&i_e&m^+\\j_{2,e-1}&m^-&i_e}}{\theta(s_e,j_{2,e},j_{2,e})}
		\frac{\Tet{s_e&j_{2,e}&i_e\\j_{1,e}&i_e&j_{2,e}}}{\theta(s_e,m^-,m^+)}
		.
\end{multline}
Hence, we can factor $M_e$ as follows:
\begin{multline}
	(M_e)^{j_{2,e} i_{e+1}}_{j_{2,e-1} i_e} =
	\frac{(i_e+1)\, \psi_{2,e-1}(j_{2,e-1})}
		{\theta(j_{2,e},i_{e+1},m^-)\,\theta(j_{2,e},i_{e+1},m^+)}
	\sum_{s_e,j_{1,e}}
		\frac{\psi_{1,e}(j_{1,e}) \Delta_{s_e}}{\theta(j_{2,e},i_e,j_{1,e})} \\
		\frac{\Tet{s_e&m^+&j_{2,e}\\i_{e+1}&j_{2,e}&m^-}}{\theta(s_e,i_e,i_e)}
		\frac{\Tet{s_e&i_e&m^+\\j_{2,e-1}&m^-&i_e}}{\theta(s_e,j_{2,e},j_{2,e})}
		\frac{\Tet{s_e&j_{2,e}&i_e\\j_{1,e}&i_e&j_{2,e}}}{\theta(s_e,m^-,m^+)}
		.
\end{multline}
Graphically, this rewriting can be show to be a factorization:
\begin{equation}\label{bcb-m7}
	M_e = \phantom{i_{e+1}} \bcbMs \phantom{j_{2,e-1}},
\end{equation}
where the factors are given explicitly by
\begin{align}
	(A_e^{j_{2,e}})^{i_{e+1}}_{s_e} &=
		\frac{\Tet{s_e&m^+&j_{2,e}\\i_{e+1}&j_{2,e}&m^-}}
			{\theta(j_{2,e},i_{e+1},m^-)\,\theta(j_{2,e},i_{e+1},m^+)}, \\
	(\psi{-}B_e^{s_e})^{j_{2,e}}_{i_e} &= \sum_{j_{1,e}}
		\frac{\psi_{1,e}(j_{1,e}) \Delta_{s_e}}{\theta(j_{2,e},i_e,j_{1,e})}
		\frac{\Tet{s_e&j_{2,e}&i_e\\j_{1,e}&i_e&j_{2,e}}}
			{\theta(s_e,j_{2,e},j_{2,e})}, \\
	(C_e^{i_e})^{s_e}_{j_{2,e-1}} &= (i_e+1)
		\frac{\Tet{s_e&i_e&m^+\\j_{2,e-1}&m^-&i_e}}
			{\theta(s_e,i_e,i_e)\,\theta(s_e,m^-,m^+)}.
\end{align}
The decomposition is not completely unique; some of the terms may be
distributed differently among the factors. However, in this
factorization, the dependence of $\psi_{1,e}$ can only be generalized to
$(j_{1,e},j_{2,e})$.

Thus $M_e$ is clearly decomposed into sparse factors, as each of $A_e$,
$B_e$ and $C_e$ is dense in some indices, but diagonal in others.
Computing the run time complexity, we get $O(j^7)$, as $d=2$, $D=2+1$,
and $2+d+D=7$, while $f=0$, $F=3$ and $F+f=3$ for filling either $A$ or
$C$.  Note that the matrices $B_e$ contracted with the $\psi_{1,e}$
factors do not depend on $m^\pm$. Hence, their computation can be done
outside the $m^\pm$ summation loops and becomes completely subdominant.

Curiously, the most practically efficient implementation of the
algorithm described in this section, as carried out by
Christensen~\cite{Ch-priv}, turns out to be a hybrid of $O(j^8)$ and
$O(j^7)$ versions. The factorization~\eqref{bcb-m7} greatly speeds up
the matrix filling step, while the simplicity of the dense version of
the product-trace operation is still advantageous for all inputs tried
to date (up to about $j_0=10$).

\subsection{New vertices with boundary states}
For the EPR and FK models, consider respectively
\begin{align}
	\label{epr-psi}
	\Psi(\{j_{c,e},i_{e}\}) &= \prod_{c,e} \psi_{c,e}(j_{c,e})
		\prod_{i_e} \psi_{e}(i_e)
\intertext{and}
	\label{fk-psi}
	\Psi(\{j_{c,e},i_{e},k^x_{c,e}\}) &= \prod_{c,e} \psi_{c,e}(j_{c,e})
		\prod_{i_e} \psi_{e}(i_e) \prod_{c,e} \psi_{x,c,e}(k^x_{c,e})
	.
\end{align}
Again, we shall only discuss the FK model explicitly, as the EPR model
can be directly obtained by dropping $k$-dependent $\psi$s and
substituting $k=j$ everywhere else.

Essentially, we want to compute the quantity $Z_\Psi$ from
equation~\eqref{bdr-contract} with a suitably factorable boundary state
$\Psi$ and the vertex amplitude specified by equation~\eqref{new-m}.
This expression for $Z_\Psi$ can be cast into the form~\eqref{CE-bdr}
with the following redefinition of $M_e$, given directly in graphical
form:
\begin{equation}
	M_e = \phantom{i_{e+1}} \newbM \phantom{j_{2,e-1}},
\end{equation}
where $\bar{T}$ denotes a product of $T$ and $N$ from
equations~\eqref{new-T} and~\eqref{new-N}. Writing out this
factorization of $M_e$ with all indices shown explicitly, while straight
forward, is cumbersome and not particularly enlightening.  It should now
be clear that, for this factorization of the $M_e$, individual factors
of the boundary state may depend on clusters of spins of the form
$(i_e,j_{1,e},j_{2,e},k^p_{1,e},k^p_{2,e})$ as well as
$(i_{e+1},j_{1,e},j_{2,e-1},k^q_{1,e},k^q_{2,e-1})$, which are
compatible with possible factorizations of the boundary states proposed
in sections~\ref{wavep} and~\ref{gprop}.

Each $M_e$ is of size $O(j^4)\times O(j^4)$, hence $d=4$. However,
because of the sparseness of the $\bar{T}$, $P$, and $Q$ factors, each
matrix-vector multiply takes $O(j^6)$ operations, since $D=4+2$ for $P$
and $Q$ multiplies and, equivalently in terms of complexity, $D=5+1$ for
each $\bar{T}$ multiply. These numbers are identical for both EPR and FK
models. On the other hand, filling the $P$ and $Q$ matrices for the EPR
model does not involve summations over $k$-spins. Thus, the EPR filling
complexity is parametrized by $f=1$, $F=5$, and $F+f=6$, while the FK
filling complexity is parametrized by $f=3$, $F=5$, and $F+f=8$.  The
overall runtime complexity of the algorithm is $O(j^{12})$, $2+d+D=12$,
both for the EPR and FK models. By conventional standards, this
algorithm has a very high polynomial complexity exponent. However, it is
still a substantial improvement over the naive $O(j^{22})$ or
$O(j^{42})$ estimates found earlier.

\section{Applications of the algorithms\label{apps}}
The algorithms described in the preceding sections have already been
implemented and applied in several contexts, other than the results
presented in this section. Alesci, Bianchi, Magliaro and
Perini~\cite{ABMP-num} have used one variation to extend the original
wave packet propagation calculations of~\cite{MPR}, both to larger input
spins and to different kinds of observables (although still keeping the
$j$-spins frozen). Also, a highly optimized version of the algorithm
presented in section~\ref{bc-bdr} has been implemented by
Christensen~\cite{Ch-priv} and used to extract next-to-leading-order
asymptotics information from the BC graviton propagator (cf.\ 
section~\ref{gprop}), as a follow-up to~\cite{CLS}. While the method
used in~\cite{CLS} is capable of handling higher input spins, the
advantage of the new algorithm is much greater precision, which is
better suited for subdominant asymptotics analysis.

Here, we apply the new algorithms to the problems of comparison of
amplitude asymptotics and of wave packet propagation, described
respectively in sections~\ref{bc-compar} and~\ref{wavep}. We have
already established that the boundary states proposed in the latter
section are factored states compatible with the new algorithms. However,
being gaussian, they do not have finite support. Fortunately, strong
gaussian decay allows us to impose a finite cutoff while maintaining
acceptable precision. The cutoff chosen for all computations presented
below was $2.8$ standard deviations about the mean. As a consequence,
the range of each spin sum involved in the computation is still of order
$O(j)$, as assumed by our run time complexity estimates.

\subsection{Amplitude asymptotics\label{ampl-res}}
As described in section~\ref{bc-compar}, because of the different spin
argument structure for each of the models under consideration, the
comparison of their amplitudes has to be done at the level of effective
vertex amplitudes given in equation~\eqref{Aeff-def}. The results of the
maximization procedure outlined previously are given below.

\begin{figure}
\begin{center}
	\includegraphics{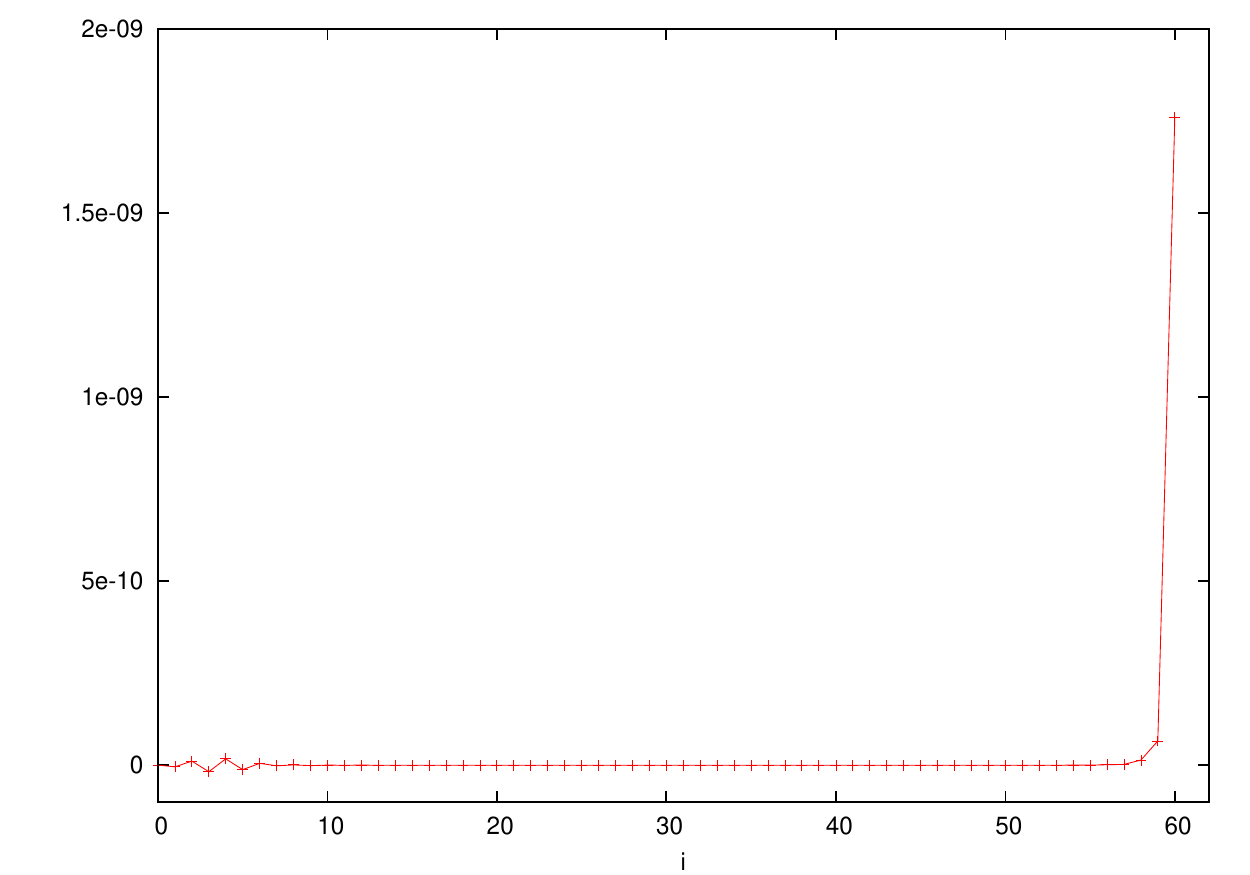}
	\caption{Effective EPR vertex amplitude: all $j=30$, all $i$ equal,
	satisfying $0\le i\le 2j$.\label{epr-fixj-vari}}
\end{center}
\end{figure}

For the EPR model, we found that the maximum allowed value $i=2j$
maximizes the amplitude. This behavior is illustrated in
figure~\ref{epr-fixj-vari} for $j=30$.

\begin{figure}
\begin{center}
	\includegraphics[scale=.96]{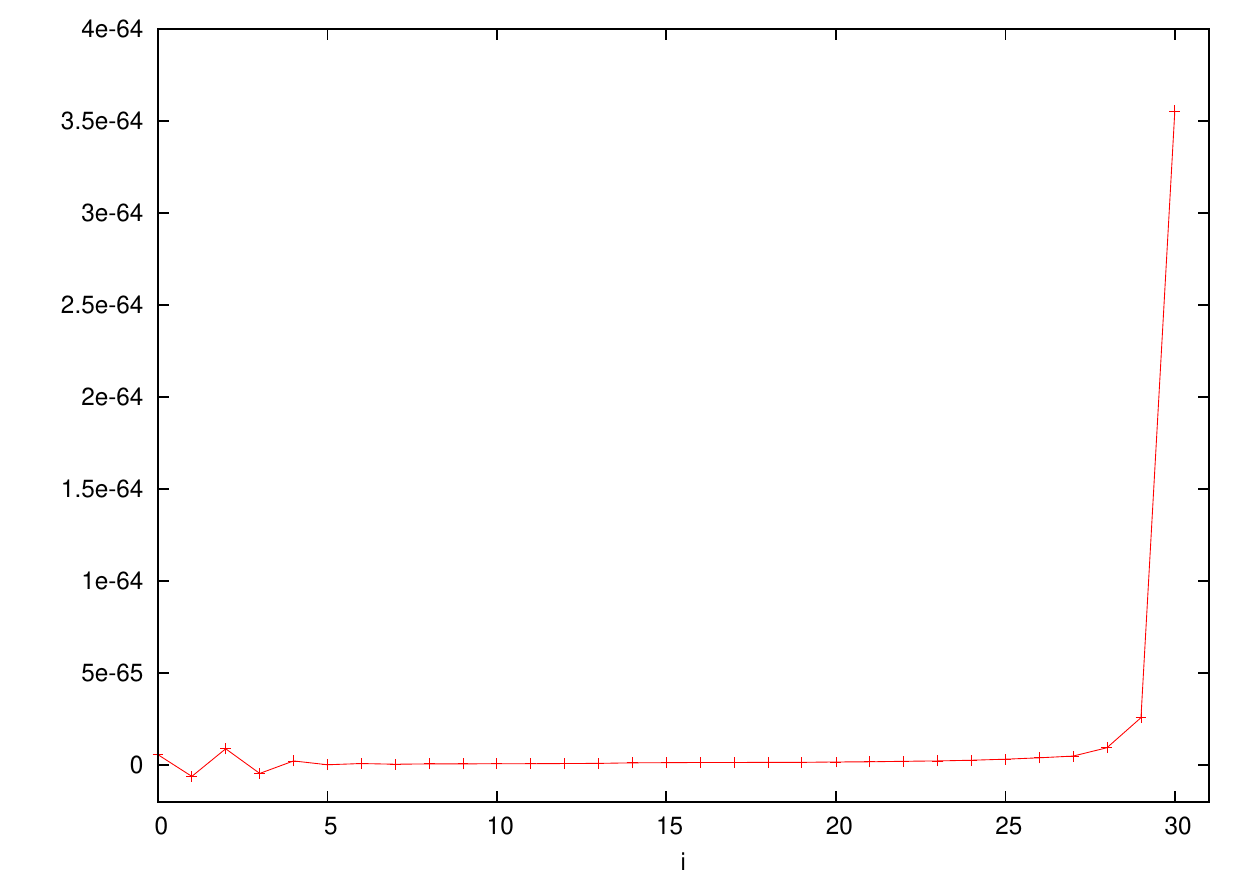}
	\caption{Effective FK vertex amplitude: all $j=30$, all $k=15$, all
	$i$ equal, satisfying $0\le i\le 2k$.\label{fk-fixj-fixk-vari}}
\end{center}
\end{figure}

\begin{figure}
\begin{center}
	\includegraphics[scale=.96]{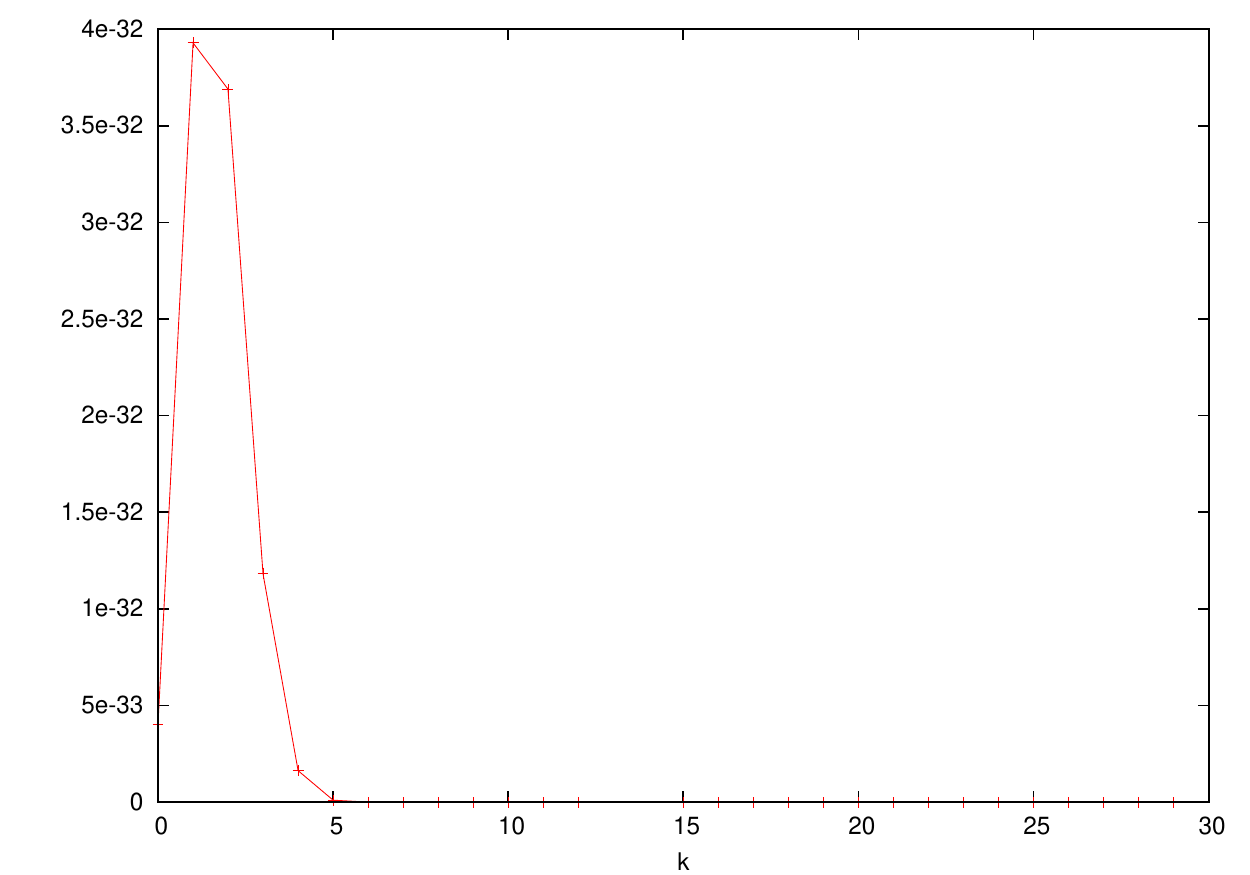}
	\caption{Effective FK vertex amplitude: all $j=30$, all $i=2k$, all
	$k$ equal, satisfying $0\le k\le j$.\label{fk-fixj-vark-fixi}}
\end{center}
\end{figure}

For the FK model, we found that, for fixed $j$ and $k$, the amplitude is
again maximized by the extreme value $i=2k$. See
figure~\ref{fk-fixj-fixk-vari} for the case $j=30$ and $k=15$. While
keeping $i$ at the dominant value $2k$, for fixed $j$, the amplitude is
maximized by $k=1$, although $k=0$ dominates slightly for very small
values of $j$. This $k$ dependence is illustrated in
figure~\ref{fk-fixj-vark-fixi} for the case $j=30$.

\begin{figure}
\begin{center}
	\includegraphics[scale=.95]{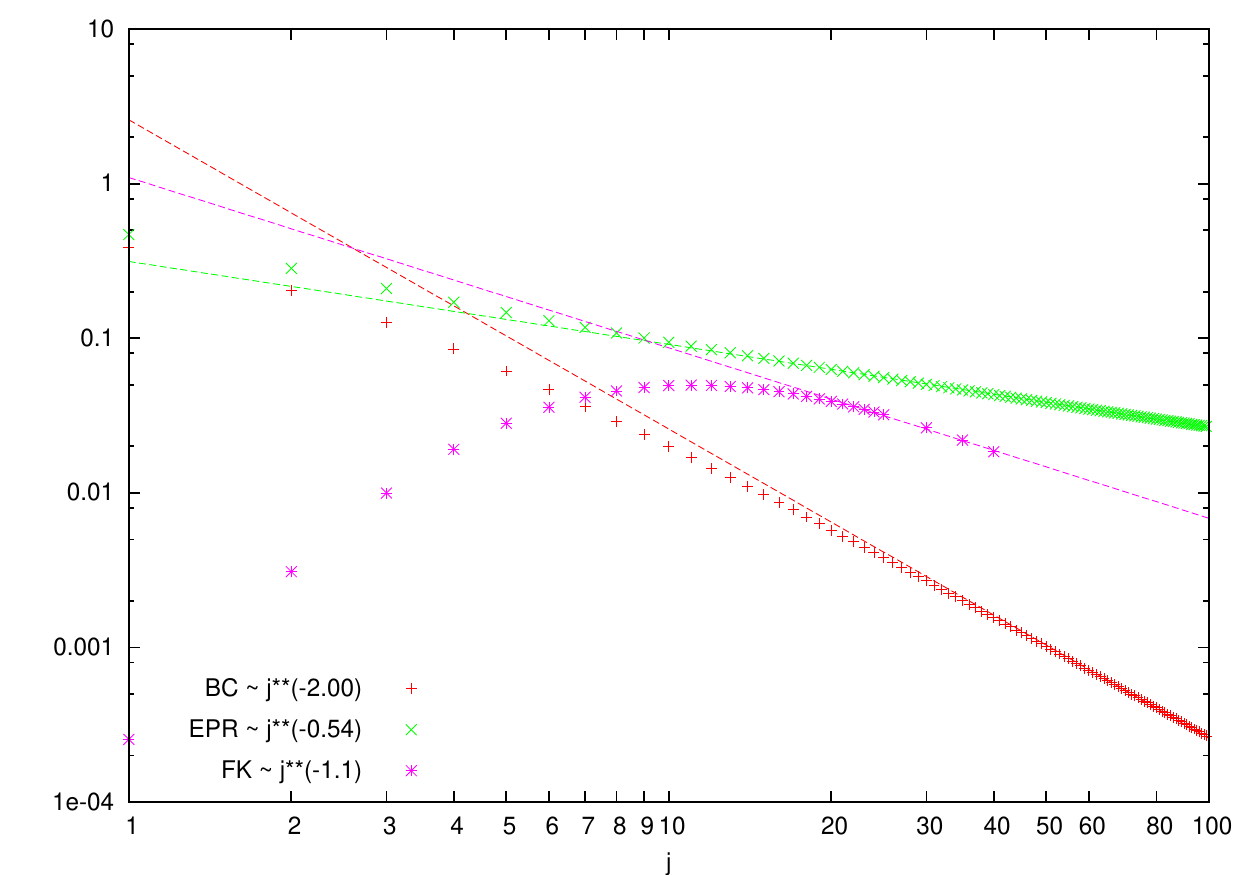}
	\caption{Large $j$ behavior of the effective vertex amplitudes for the
	BC, EPR and FK models.\label{large-j}}
\end{center}
\end{figure}

Spin foam quantization is similar in spirit to the discretized path
integral approach to gravity. As such, the spin foam vertex amplitude is
often compared to the gravitational path integral amplitude:
\begin{equation}
	A_v(j) \sim \exp[iS_R],
\end{equation}
where $S_R$ is the Regge action for gravity evaluated on a discrete
geometry described by the spins $j$ in the large spin limit.  For the BC
vertex, this view has turned out to be overly simplistic. The relation
predicted by careful asymptotic analysis is
\begin{equation}\label{asymp-struct}
	A_v(j) \sim D(j) + \mu(j) [\exp(iS_R) + \exp(-iS_R)] + \cdots ,
\end{equation}
where $D(j)$ and $\mu(j)$ are non-oscillating functions decaying as
$j^{-2}$ and $j^{-9/2}$ respectively, with $(\cdots)$ representing
higher order terms. The dominant asymptotic $D(j)$, understood to be due
to the contribution of degenerate geometries, masks the desired Regge
action amplitude~\cites{BCE,FL,BS}.

A natural question is whether the new vertices share the same asymptotic
behavior. Numerical evaluation of the BC vertex is only sensitive to the
dominant asymptotic contribution $D(j)$. The subdominant oscillating
Regge action term would become important only if $D(j)$ is subtracted or
if the vertex amplitude is averaged against another oscillatory
function, in phase with one of the Regge action terms, as done in the
graviton propagator calculations~\cites{R-prop,CLS}. While analytical
asymptotics for the EPR and FK vertices are still missing, we can
straightforwardly compare the numerical asymptotics of the dominant
effective vertex amplitudes of all three models. This comparison is made
in figure~\ref{large-j}. For all models, the data shows no oscillations,
which means that we are most likely seeing only the $D(j)$ asymptotic
term. Note that the power laws shown in the figure will change if the
edge or face amplitudes given in section~\ref{bf-th} are modified by
$j$-dependent factors.  Such modifications can come, for instance, from
different choices of exponent $k$ for factors of the form $(j_f+1)^k$,
contributing to face amplitudes, as was considered in equation~(3)
of~\cite{CLS}. Different choices of this $k$ would correspond to
different choices in the path integral analog of spin foams.

\subsection{Wave packet propagation\label{wavep-res}}
\begin{figure}
\begin{center}
	\includegraphics[width=\textwidth]{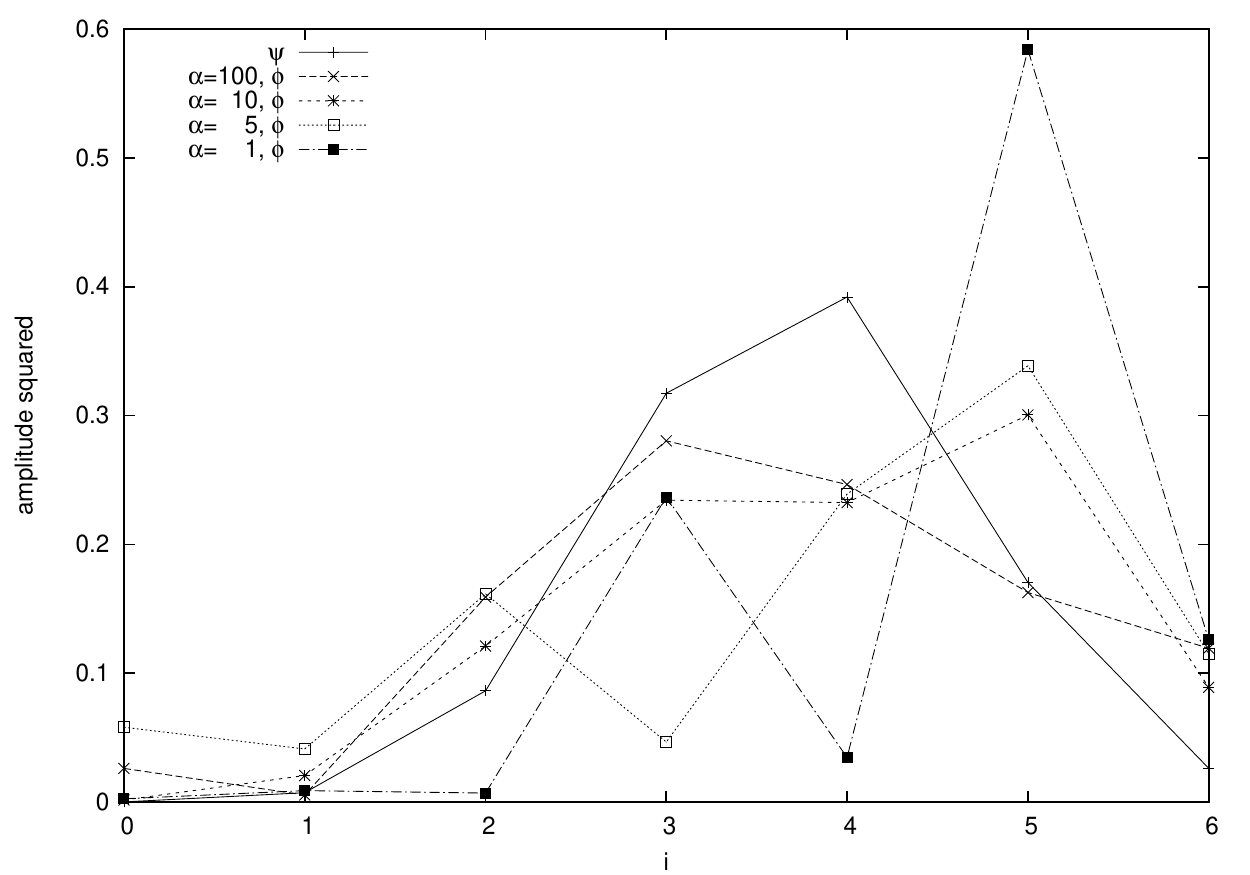}
	\caption{EPR 4-1 propagated ($\phi$) and reference ($\psi)$
	wave packets, with $j_0=3$.
	\label{epr4-1}}
\end{center}
\end{figure}

As an immediate improvement over previous work, the algorithms presented
in this paper allow us to show the effect of introducing a non-zero
$\tau$ in~\eqref{bc-gauss} and to compare with the calculations
of~\cite{MPR}, which kept $\tau=0$, freezing all $j$-spins at the
background value $j_0$. According to equation~\eqref{tau-def}, the size
of $\tau$ is inversely proportional to the parameter $\alpha$.
Figure~\ref{epr4-1} compares the reference wave packet $\psi$
[cf.~\eqref{epr-gauss}] with several propagated wave packets $\phi$
(each with a different value of $\alpha$) depending on the single fixed
$i$-spin. The wave packets have been normalized such that their absolute
values squared sum to $1$.  The wave packet with the largest value of
$\alpha$ is essentially identical to the one obtained with all $j$-spins
frozen at $j_0$. In that case, as shown previously in~\cite{MPR}, the
reference state $\psi$ resembles the propagated wave packet in shape and
mean. Unfortunately, as the width of the gaussian factors associated to
$j$-spins increases ($\alpha$ decreases), the propagated wave packet
quickly departs from $\psi$ in both shape and mean. Notably, the mean
shifts to a significantly higher value of $i$.

\begin{figure}
	\includegraphics[width=\textwidth,trim=0 30 0 30]{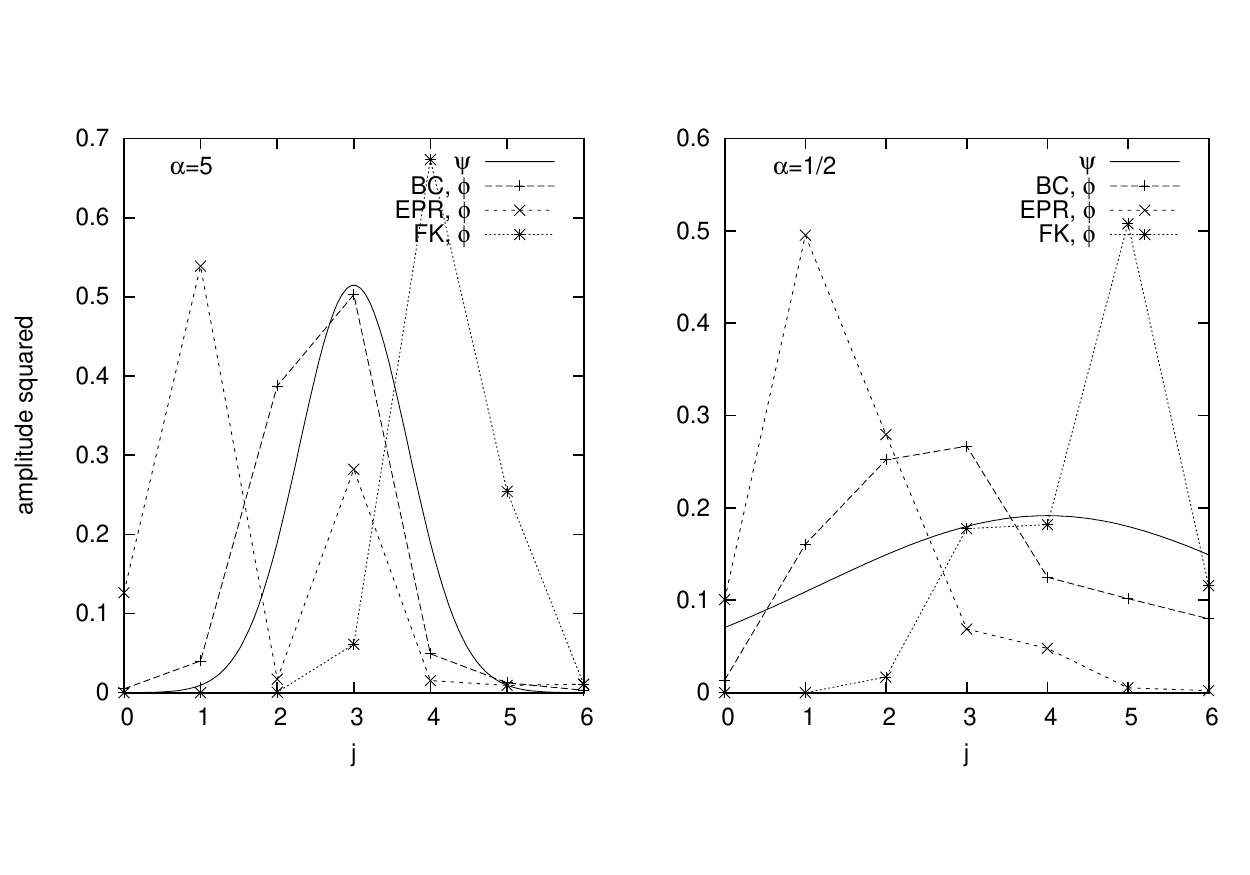}
	\caption{9-1 propagated and reference wave packets for different
	models, with $j_0=3$.
	\label{all9-1}}
\end{figure}

Second, we can compare the wave packets propagated by the three
different models in the 9-1 geometry. Figure~\ref{all9-1} shows the
reference wave packets $\psi$ [cf.~\eqref{bc-gauss}] and propagated wave
packets $\phi$, depending on the fixed $j$-spin and for two choices of
$\alpha$. These wave packets are also normalized.  While he propagated
wave packets seem to retain a roughly peaked shape, their mean and width
mostly differ significantly from the reference wave packet. The only
exception is the BC model at $\alpha=5$. However, this is likely a
coincidence that does not appear for other choices of parameter values.

\begin{figure}
	\includegraphics[width=\textwidth,trim=0 30 0 30]{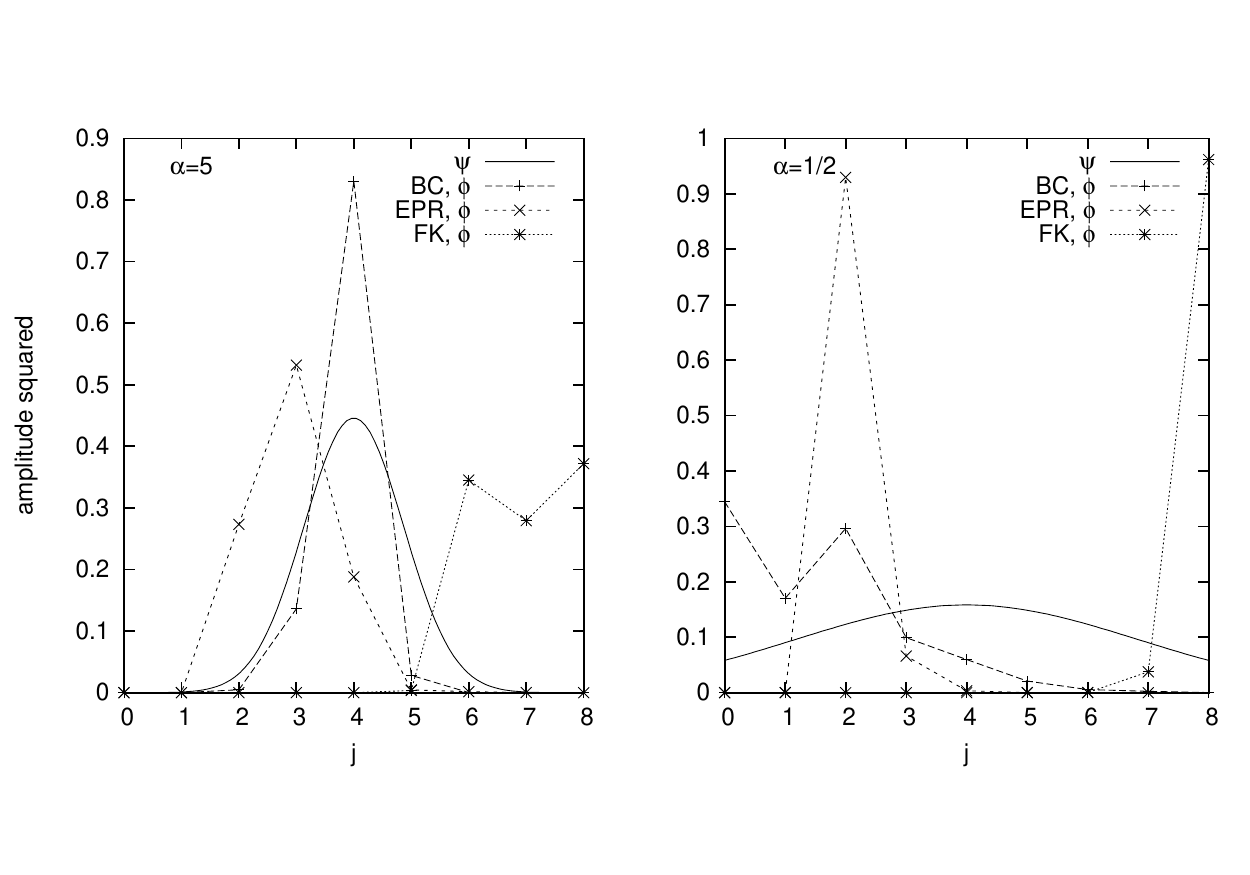}
	\caption{4-6 propagated and reference wave packets for different
	models, with $j_0=4$.
	\label{all4-6}}
\end{figure}

Lastly, we compare the wave packets propagated by the three different
models in the 4-6 geometry. In general, the propagated wave packet will
depend on the four fixed $j$-spins. Unfortunately, it is impractical to
either compute or display functions on a $4$-dimensional domain. Thus,
all calculations have been done with the four fixed $j$-spins set equal.
Figure~\ref{all4-6} shows the reference wave packet $\psi$
[cf.~\eqref{bc-gauss}] and propagated wave packets $\phi$, depending on
the common value of the fixed $j$-spins for two choices of $\alpha$.
These wave packets are again normalized. As is clearly seen from the
figure, the propagated wave packets have in general very little
similarity with the reference one. More pathological behavior is
observed in the FK and BC models, the latter at $\alpha=1/2$, since none
of these curves resemble a well formed gaussian wave packet. In all
cases, the propagated wave packet has little in common with the
reference one.

\section{Conclusion and outlook\label{concl}}
We have presented three spin foam models in a unified framework: the
standard Barrett-Crane (BC) model and two more recent proposals, the
Engle-Pereira-Rovelli (EPR) and Freidel-Krasnov (FK) models. Their
vertex amplitudes were simplified and explicitly evaluated using spin
network recoupling techniques. Despite the different spin argument
structure, we have proposed uniform methodologies for comparing the
results of calculations of asymptotics of effective vertex amplitudes
and of expectation values of specific spin foam observables among the different
models. Building on the past work of Christensen~\& Egan, a fast
numerical evaluation algorithm for the new vertex amplitudes, both in
the absence and in the presence of (factored) boundary states, was
developed and implemented. The run time complexity of these algorithms
has been analyzed and shown to be orders of magnitude superior to more
naive approaches. These algorithms were applied to the problems of
extracting the asymptotic behavior of the new vertex amplitudes and to
checking the behavior of propagated semiclassical wave packets.

The results presented in section~\ref{ampl-res} show that the dominant
asymptotic behavior of the new vertex amplitudes is non-oscillatory and
displays power-law decay very similar to the BC model, suggesting
dominance of degenerate spin foams as for the BC model itself. The power
law exponents are estimated in figure~\ref{large-j}. It should be
interesting to explore the asymptotics of these amplitudes with
analytical techniques as well. It is likely that they will reveal
structure similar to~\eqref{asymp-struct}.

Although the boundary state proposed in section~\ref{sf-bdr} is not
ideal (see~\cite{RS} for a more realistic proposal), it has the
advantage of belonging to the class of factored states, allowing
efficient numerical computation with algorithms of
section~\ref{num-alg}. Moreover, the proposed state can still be used to
gauge the qualitative behavior of unfreezing the $j$-spins.
Reference~\cite{MPR} had put forward the conjecture that semiclassical
wave packets, propagated using the EPR dual vertex amplitude,
approximate a certain reference gaussian shape, which was demonstrated
under somewhat restrictive conditions. Application of the numerical
algorithms described above allowed a broader investigation of this
question.  The class of factored boundary states encompasses the states
used in the calculations that gave rise to this conjecture. The results
of section~\ref{apps} indicate that a generic factored boundary state
does not exhibit good semiclassical behavior. It is likely that the
presence of correlations between spins, which are not not captured by a
factored state, improves the agreement between propagated and reference
wave packets. However, until the use of more complicated states is
possible, the evidence for the conjecture of Magliaro, Perini \& Rovelli
remains inconclusive.

These algorithms have also already been implemented and applied by other
authors, as discussed in section~\ref{wavep-res}. While, several wave
packet propagation geometries have been examined, there are many other
ones.  Important questions remain: Is any one of them theoretically
preferable to the others?  What is the impact of directly using the
unfactorable boundary state of~\cite{RS}? Another immediate possibility
for further investigation is the computation of the graviton propagator
matrix elements in the EPR and FK models.

\section*{Acknowledgments}
The author would like to thank Dan Christensen for suggesting this
project. Also, Wade Cherrington, Carlo Rovelli, Elena Magliaro, Claudio
Perini, Simone Speziale and Laurent Freidel have
contributed through helpful discussions. The author was supported by an
Ontario Graduate Scholarship and a SHARCNET Fellowship. Computational
resources for this project were provided by SHARCNET.

\end{document}